\documentclass{aa}
\usepackage{graphicx,aabib,amssymb}
\begin{document}
\title{Origin of quasi--periodic shells in dust forming AGB winds}
\author{Y.J.W. Simis\inst{1}, V. Icke\inst{1} and 
	C. Dominik\inst{2}}
\offprints{Yvonne Simis}
\mail{simis@strw.leidenuniv.nl}
\institute{ 
        Leiden Observatory, P.O. Box 9513, 2300 RA Leiden, The Netherlands
        \and
	Astronomical Institute ``Anton Pannekoek'', University of Amsterdam,
	Kruislaan 403, 1098 SJ Amsterdam, The Netherlands
 }

\date{Received ; Accepted}
\titlerunning{Origin of quasi--periodic shells in dust forming AGB winds}
\authorrunning{Y.J.W. Simis et al.}
%
%----------------------------------------------------------------------------
%
\abstract{ We have combined time dependent hydrodynamics with a
two--fluid model for dust driven AGB winds. Our calculations include
self--consistent gas chemistry, grain formation and growth, and a new
implementation of the viscous momentum transfer between grains and
gas. This allows us to perform calculations in which no assumptions
about the completeness of momentum coupling are made. We derive new
expressions to treat time dependent and non--equilibrium drift in a
hydro code.  U\-sing a stationary state calculation for \object{IRC
+10216} as initial model, the time dependent integration leads to a
quasi--periodic mass loss in the case where dust drift is taken into
account. The time scale of the variation is of the order of a few
hundred years, which corresponds to the time scale needed to explain
the shell structure of the envelope of \object{IRC +10216} and other
AGB and post-AGB stars, which has been a puzzle since its
discovery. No such periodicity is observed in comparison models
without drift between dust and gas.
\keywords{Hydrodynamics -- Methods: numerical -- Stars: AGB and
post-AGB -- Stars: mass loss -- Stars: winds, outflows -- Stars:
individual: IRC +10216} } 
\maketitle
%
%----------------------------------------------------------------------------
%
\section{Introduction}
Dust driven winds are powered by a fascinating interplay of radiation,
chemical reactions, stellar pulsations and dynamics. As soon as the
envelope of a star on the Asymptotic Giant Branch (AGB) develops sites
suitable for the formation of solid ``dust'' (i.e. sites with a
relatively high density and a low temperature) its dynamics will be
dominated by radiation pressure. Dust grains are extremely sensitive
to the stellar radiation and experience a large radiation
pressure. The acquired momentum is partially transferred to the
ambient gas by frequent collisions.  The gas is then blown outward in
a dense, slow wind that can reach high mass loss rates.\\
The detailed observations of (post) AGB objects and Planetary Nebulae
(PN) that have become available during the last decade have shown that
winds from late type stars are far from being smooth. The shell
structures found around e.g. \object{CRL 2688} (the ``Egg Nebula'',
\cite{Ney:1975}; \cite{Sahaiea:1998}), \object{NGC 6543} (the Cat's
Eye Nebula, \cite{HarBork:1994}) and the AGB star \object{IRC +10216}
(Mauron \& Huggins 1999, 2000), indicate that the outflow has
quasi--periodic oscillations. The time scale for these oscillations is
typically a few hundred years, i.e. too long to be a result of stellar
pulsation, which has a period of a few hundred days, and too short to
be due to nuclear thermal pulses, which occur once in ten thousand to
hundred thousand years.\\
Stationary models, in which gas and dust move outward as a single
fluid, do not suffice to explain the observations. Instead, time
dependent two--fluid hydrodynamics, preferably including (grain)
chemistry and radiative transfer, may help to explain the origin of
these circumstellar structures.\\
Time dependent hydrodynamics has been used to study the influence of
stellar pulsations on the outflow
(\cite{Bowen:1988,Fleischerea:1992}). The coupled system of radiation
hydrodynamics and time dependent dust formation was solved by
\cite*{Hoefnerea:1995}.\\
Stationary calculations, focused on a realistic implementation of
grain nucleation and growth, have been developed in the Berlin group,
initially for carbon--rich objects
(\cite{GailKellerSedlmayr:1984,GailSedl:1987}) and more recently also
for the more complicated case of silicates in circumstellar shells of
M stars (\cite{GailSedl:1999}).\\
Two--fluid models, in which dust and gas are not necessarily
co--moving, have been less well studied.  \cite*{BerruFrisch:1983},
\cite*{Berruyer:1991} and \cite*{MacGregSten:1992}, pointed out that,
for stationary and isothermal envelopes, the assumption of complete
momentum coupling breaks down at large distances above the photosphere
and for small grains. Self--consistent, but again stationary,
two--fluid models, considering the grain size distribution, dust
formation and the radiation field were developed by Kr\"uger and
co--workers (\cite{Kruegerea:1994}; \cite{KruegSedl:1997}).\\
The only studies in which time dependent hydrodynamics and two--fluid
flow have been combined so far are the work of
\cite*{Mastrodemosea:1996} and that of the Potsdam group
(\cite{Steffenea:1997}; \cite{Steffenea:1998};
\cite{SteffenSchoenberner:2000}).\\
In the next section, we will argue that time dependence and two--fluid
flow are not just two interesting aspects of stellar outflow but that
they have to be combined. It turns out that fully free two--fluid
flow, i.e. in which no assumptions at all about the amount of momentum
transfer between both phases are made, can only be achieved in time
dependent calculations.  In two--fluid flow, both phases are described
by their own continuity and momentum equations. Momentum exchange
occurs through viscous drag, i.e. through gas--grain collisions. The
collision rate and the momentum exchange per collision depend on the
velocity of grains relative to the gas. Hence, by fixing the drag
force, one fixes the relative velocity and the system becomes
degenerate.\\
In this paper we present our two--fluid time dependent hydrodynamics
code.  We have selfconsistently included equilibrium gas chemistry and
grain nucleation and growth, see Section \ref{sec:modelmethod}.  In
order not to make assumptions on the viscous coupling, we consider, in
Section \ref{sec:momcoup}, the microphysics of gas--grain collisions.
Results are given in Section \ref{sec:results}.
%
%----------------------------------------------------------------------------
%
\section{Grain drift and momentum coupling}
\label{sec:driftandcoupling}
\subsection{Definitions}
\label{sec:definitions}
The acceleration of dust grains, as a result of radiation pressure,
leads to an increase in the gas--dust collision rate. The viscous drag
force (the rate of momentum transfer from grains to gas due to these
collisions) is proportional to the collision rate and to the relative
velocity of grains with respect to the gas. This force is discussed in
the next section in more detail. The drag force provides a (momentum)
coupling between the gaseous and the solid phase\footnote{Another
momentum coupling is due to the fact that momentum is removed from the
gas phase when molecules condense on dust grains. The amount of
momentum involved in this coupling is also taken into account in our
numerical models but is many orders of magnitude smaller than the
collisional coupling.}.\\
The gas--dust coupling was studied by e.g. \cite*{Gilman:1972}, who
distinguished two types of coupling.  Gas and grains are {\em position
coupled} when the difference in their flow velocities, the drift
velocity, is small compared to the gas velocity, i.e. when the grains
move slowly through the gas. {\em Momentum coupling}, on the other
hand, requires that the momentum acquired by the grains through
radiation pressure is approximately equal to the momentum transferred
from the grains to the gas by collisions.  The situation in which both
are exactly equal is called {\em full} or {\em complete} momentum
coupling. \cite*{Gilman:1972} stated that, if both forces are equal,
grains drift at the {\em terminal} drift velocity. A less confusing
term for the same situation was introduced by
\cite*{Dominikthesis:1992}: {\em equilibrium drift}. The idea is that
since the drag force increases with increasing drift velocity, an
equilibrium value can be found by equating the radiative acceleration
of the grains and the deceleration due to momentum transfer to the
gas. Note that, when calculating the equilibrium value of the drift
velocity that way, i.e. assuming complete momentum coupling, one
implicitly assumes that grains are massless. A physically correct way
to calculate the equilibrium drift velocity is to demand gas and
grains to have the same acceleration.
\subsection{Single and multi--fluid models}
\label{sec:singmultimodels}
Various groups have studied the validity of momentum coupling, with
and without assuming equilibrium drift, in stationary and in time
dependent calculations. Others have just applied a certain degree of
momentum coupling in model calculations carried out to study other
aspects of the wind. We will give a brief overview of the most
important of these studies, resulting in the conclusion that prior to
our attempt, full two--fluid hydrodynamics has been presented only
twice. Because the meaning of terms like ``full'' and ``complete''
momentum coupling, ``terminal'' and ``equilibrium'' drift seem to be
slightly different from author to author, we will first give our own
definitions for three classes of models.\\
First, {\em single--fluid} models are those in which only the momentum
equation of the gas component is solved. All momentum due to radiation
pressure on grains is transferred fully and instantaneously to the
gas. If, e.g., for the calculation of grain nucleation and growth
rates, a value for the flow velocity of the dust component is needed,
the dust is just assumed to have the same velocity as the gas: drift
is assumed to be negligible.  Hence, in terms of \cite*{Gilman:1972},
in single fluid models grains are both position and (completely)
momentum coupled to the gas.\\
The second class is that of the {\em two--fluid} models.  Here, again
in terms of \cite*{Gilman:1972}, grains are not necessarily position
and momentum coupled to the gas. Grains can drift at non--equilibrium
drift velocities. Hence, grains and gas are neither forced to have
equal velocity nor forced to have equal acceleration.\\
The third category of models represents what we will call {\em
1.5--fluid} models. In these models, grains are assumed to drift at
the equilibrium drift velocity with respect to the gas. No assumptions
about position coupling are made. In other words, gas and grains are
equally accelerated but do not necessarily have the same velocity. The
equilibrium drift velocity is calculated by equating the drag force
and the radiation pressure on the grains, see
\cite*{Dominikthesis:1992}, or, more accurately, by demanding gas and
grains to be equally accelerated. Only the momentum equation of the
gas is solved, the dust velocity is determined by simply adding the
gas velocity and the equilibrium drift velocity.
\subsection{Stationary models}
\label{sec:stationarymodels}
Although the above classification for modeling methods also applies to
stationary models, extra care is needed there. When trying to do
two--fluid stationary modeling one should realize that the condition
of stationarity {\em itself} will also introduce momentum coupling.
This can be understood as follows. Equilibrium drift is the state in
which gas and grains are equally accelerated:
\begin{equation}
\frac{\mathrm{d} v_\mathrm{g}}{\mathrm{d} t} = \frac{\mathrm{d} v_\mathrm{d}}{\mathrm{d} t}
\label{eq:dvgdtisdvddt}
\end{equation}
The derivative in this equation is a total derivative.  Imposing
stationarity, the temporal contribution to this total derivative
vanishes by definition, and Eq.(\ref{eq:dvgdtisdvddt}) reduces to
\begin{equation}
v_\mathrm{g}\frac{\partial v_\mathrm{g}}{\partial r} = 
v_\mathrm{d}\frac{\partial v_\mathrm{d}}{\partial r}
\label{eq:dvgdrisdvddr}
\end{equation}
The difference between both sides of Eq.(\ref{eq:dvgdrisdvddr}) can
be small, especially in the outer layers of the envelope, where the
velocities reach a more or less constant value.  Therefore, the
occurrence of equilibrium drift in a stationary outflow may be
partially due to the condition of stationarity itself.  For this
reason, one should be very careful when checking the validity of
momentum coupling against stationary calculations.  Moreover, in order
to make a calculation fully self--consistent, no assumptions on
momentum coupling should be made. Hence, for fully self--consistent
modeling, time dependent calculations are to be preferred.
\subsection{Overview of previous modeling}
\label{sec:previousmodeling}
Examples of single fluid calculations are naturally found in studies
in which drift and momentum coupling are not the topic of research,
e.g. the work of \cite*{DorfiHoefner:1991} and
\cite*{Fleischerea:1995}. Both perform time dependent hydrodynamics,
assuming that the influence of drift on the aspect of the flow under
consideration, dust formation and nonlinear effects due to dust
opacity, is negligible.\\
The completeness of momentum coupling is investigated by
\cite*{BerruFrisch:1983} and by \cite*{Kruegerea:1994}.  The former
first find a (stationary) wind solution under the assumption of
complete momentum coupling, noticing that this assumption causes the
two--fluid character to be lost. Next, in order to check the validity
of their supposition, they find a stationary solution for the system,
including the grain momentum equation. Both calculations give very
similar results near the photosphere, from which it is concluded that
momentum coupling is complete there. Far away from the stellar surface
($\gtrsim 1000 R_\mathrm{*}$), the results are different so that
momentum coupling is said to be invalid there. We too, find that
non--equilibrium drift arises far away from the photosphere (see
Section \ref{sec:results}).  We would like to remark, however, that it
may not be sufficient to verify the validity of complete momentum
coupling by comparing with stationary calculations, see Section
\ref{sec:stationarymodels}.\\
\cite*{Kruegerea:1994} undertook a similar study, which is the most
realistic stationary two--fluid calculation up to now.  It treats the
coupled system of hydrodynamics and thermodynamics, but also involves
chemistry and dust formation (simplified by the assumption of
instantaneous grain formation). Kr\"uger et al. conclude that momentum
coupling can be assumed to be complete and therefore disagree with
\cite*{BerruFrisch:1983}.  We think this may be due to the fact that
Kr\"uger et al. run their calculation out to about ten stellar radii,
whereas Berruyer \& Frisch compute outwards to several thousand
stellar radii.\\
According to \cite*{MacGregSten:1992}, who use a simple model for
grain growth in a stationary, isothermal atmosphere, the assumption of
complete momentum coupling appears to break down for grain sizes
smaller than about $5 \times 10^{-6}$ cm. \\
Prior to our attempt, time dependent two--fluid hydrodynamics was
presented by \cite*{Mastrodemosea:1996}. They conclude that
fluctuations on the time scale of the variability periods of Miras and
LPV (Long Period Variables), 200-2000 days, can not persist in the
wind. Since they do not calculate grain nucleation and growth
self--consistently but instead assume that grains grow instantaneously
and have a fixed size, the extreme non--linear coupling between shell
dynamics, chemistry and radiative transfer (cf. \cite{sedlmayr:1997})
is not present. Our calculations however indicate that this
chemo--dynamical coupling is a main ingredient to the occurence of
variability in the wind.\\
Steffen and co--workers (\cite{Steffenea:1997}; \cite{Steffenea:1998};
\cite{SteffenSchoenberner:2000}) have a more or less similar approach:
their models are based on time dependent, two--fluid radiation
hydrodynamics and grains have a fixed size.  Main emphasis is on the
long term variations of stellar parameters ($L_{*}(t), \dot{M}(t)$),
due to the nuclear thermal pulses, which are included as a time
dependent inner boundary. It turns out that these large--amplitude
variability at the inner boundary is not damped in the envelope and
remains visible in the outflow as a pronounced shell.\\
The calculations presented in this paper aim at combining time
dependent hydrodynamics with a two--fluid model and are suitable for
calculating the stellar wind from the subsonic photosphere to the
supersonic outer layers at large distances. We will not take stellar
pulsation into account because we want to find out if the envelope
itself possesses characteristic time scales.  The main goal of this
work is to get insight in the physical processes underlying the
observed time dependent structures around AGB stars.  We do not aim at
exactly reproducing certain observational results and hence will not
adjust the stellar parameters in order to provide a better fit.
%
%----------------------------------------------------------------------------
%
\section{Modeling method}
\label{sec:modelmethod}
\subsection{Basic equations}
The basic equations for the time dependent description of a stellar
wind in spherical coordinates and symmetry, are the continuity
equations,
\begin{equation}
\frac{\partial \rho_\mathrm{g,d}}{\partial t} + 
\frac{1}{r^{2}}\frac{\partial}{\partial r} 
(r^{2}\rho_\mathrm{g,d}v_\mathrm{g,d}) = s_\mathrm{cond,g,d} 
\label{eq:contgd}
\end{equation}
and the momentum equations,
\begin{eqnarray}
\frac{\partial}{\partial t}(\rho_\mathrm{g}v_\mathrm{g}) &+&  
\frac{1}{r^{2}}\frac{\partial}{\partial r}
(r^{2}\rho_\mathrm{g}v_\mathrm{g}^{2}) = \nonumber \\ 
&-& \frac{\partial P}{\partial r} + f_\mathrm{drag,g}
- f_\mathrm{grav,g} + v_\mathrm{g}s_\mathrm{cond,g}
\label{eq:momg}\\
\frac{\partial}{\partial t}(\rho_\mathrm{d}v_\mathrm{d}) &+&  
\frac{1}{r^{2}}\frac{\partial}{\partial r}
(r^{2}\rho_\mathrm{d}v_\mathrm{d}^{2}) = \nonumber \\ 
&&f_\mathrm{rad} + f_\mathrm{drag,d} - f_\mathrm{grav,d}
- v_\mathrm{g}s_\mathrm{cond,g}
\label{eq:momd}
\end{eqnarray} 
These equations form a system in which both gas and dust are described
by their own set of hydro equations (two--fluid hydrodynamics). The
equations are coupled via the source terms.  The source term in
Eq.(\ref{eq:contgd}) represents the condensation of dust from the gas,
including nucleation and growth. Since mass is conserved we have
\begin{equation}
s_\mathrm{cond,g} = -s_\mathrm{cond,d}
\end{equation}
The gas condensation source term is negative due to nucleation and/or
growth of grains. Atoms and molecules that condens onto grains take
away momentum from the gas. This is accounted for in the
$v_\mathrm{g}s_\mathrm{cond,g}$ source terms in the momentum
equations.\\
The momentum equations also couple via the viscous drag force of
radiatively accelerated dust grains on the gas. Since no momentum is
lost, we have
\begin{equation}
f_\mathrm{drag,g} = -f_\mathrm{drag,d}
\end{equation}
The drag force is proportional to the rate of gas--grain collisions
and the momentum exchange per collision and is therefore of the form
\begin{equation}
f_\mathrm{drag} = \Sigma_\mathrm{d} n_\mathrm{g} n_\mathrm{d} m_\mathrm{g} |v_\mathrm{D}| v_\mathrm{D}
\label{eq:fdragsimple}
\end{equation}
where $\Sigma_\mathrm{d}$ is the collisional cross section of a dust
grain and $v_\mathrm{D}$ is the drift velocity of the grains with
respect to the gas. \\
We assume a grey dust opacity and take the extinction cross section of
the grains equal to the geometrical cross section. Then the radiative
force is simply
\begin{equation}
f_\mathrm{rad} = \frac{L_{*} \Sigma_\mathrm{d} n_\mathrm{d}}{4 \pi r^2 c}
\end{equation}
Radiation pressure on gas molecules is negligible in the circumstellar
environment of AGB stars.  In order to determine the temperature
structure of the envelope, a balance equation for the energy can be
added. We do not involve the energy structure in the time dependent
calculation. Also, we do not solve radiation transport. Instead, we
assume that, throughout the envelope, the temperature stratification
is determined by radiation equilibrium of the gas. This assumption is
justified as long as the envelope is optically thin to the cooling
radiation emitted by the dust. The inclusion of an energy equation
poses no problems, if one wants to spend the computer time.\\
The model is completed with the equation of state for ideal gases.\\
\subsection{Gas chemistry}
Our hydrocode contains an equilibrium chemistry module
(\cite{Dominikthesis:1992}) which includes H, H$_2$, C, C$_2$, C$_2$H,
C$_2$H$_2$ and CO, and hence is suitable for modeling C stars.\\
Oxygen has completely associated with carbon to form CO.  Due to the
high bond energy of the CO molecule (11.1 eV), this molecule is the
first to form.  In absence of dissociating UV radiation, CO--formation
is irreversible. Hence if $\epsilon_C > \epsilon_O$ at the time of CO
formation, all oxygen will be captured in CO and carbon will be
available for the formation of molecules and dust.  Given the total
number density of H and C atoms in the gas phase, the dissociation
equilibrium calculation is carried out in each numerical time step to
give the densities of the molecules mentioned. Therefore, bookkeeping
of the H and C number densities is needed.  This requires two
additional continuity equations of the form of Eq.(\ref{eq:contgd}).
\subsection{Grain nucleation and growth}
Once the abundances of the gas molecules are known, the nucleation and
growth of dust grains can be calculated. We use the moment method
(\cite{GailKellerSedlmayr:1984,GailSedl:1988}), in conservation form
(\cite{DorfiHoefner:1991}). The resulting nucleation and growth rates
are used to calculate the source terms of Eq.(\ref{eq:contgd}) and
the additional continuity equations for hydrogen and carbon. The
moment equations provide the evolution in time of the zeroth to third
moment of the grain size distribution function. Hence, amongst others,
the number density and the average grain size are known as a function
of time.  We could, in principle, calculate the full grain size
spectrum, using the moment method, but we limit ourselves to the use
of average grain sizes. The main advantage of this is that we can
apply two--fluid, instead of multi--fluid hydrodynamics, which is
obviously computationally cheaper.
\subsection{Viscous gas--grain momentum coupling}
\label{sec:momcoup}
In the absence of grain drift, gas and dust particles will collide
frequently due to the thermal motion of the gas, but no net momentum
transfer from one state to the other will take place since the
collisions are random. If grains are radiatively accelerated with
respect to the gas, both the thermal motion and the acceleration give
rise to gas--grain encounters, resulting in a net momentum transfer
from grains to gas. The resulting viscous drag force is described in
e.g. \cite*{Schaaf:1963}.\\
In the hydrodynamical regime, the time scale on which individual
gas--grain collisions occur is many orders of magnitude smaller than
the dynamical time scale. Hence, in order to calculate the momentum
transfer from grains to gas, one needs to sum over many collisional
events. The strong dependence of the momentum source term on the
(drift) velocity, via the drag force (Eq.(\ref{eq:fdragsimple})),
enables rapid changes in the velocities. When applying an explicit
numerical difference scheme, as we do, it will therefore be necessary
to take small numerical time steps. Taking small, and hence more, time
steps involves the risk of losing accuracy however. In our case, the
drag force makes the system so stiff that this would lead to
unacceptably small numerical time steps: a reduction of a factor
thousand or more, compared to the Courant timestep is not unusual.  To
avoid having to take such small steps we perform a kind of subgrid
calculation for the drift velocity by studying the microdynamics of
the gas--grain system. Doing so, we derive an expression for the
temporal evolution of the drift velocity during one numerical time
step. This expression is then used to calculate an accurate value of
the momentum transfer, i.e. the integrated drag force, in one
numerical time step. This way, the momentum transfer rate is
determined without making assumptions about the value of the drift
velocity at the end of the numerical time step. Hence, if the momentum
transfer is determined in this manner a full two-fluid calculation can
be done.  Details of the derivation are given in Appendix \ref{sec:append}.\\
Another way to go around the problem of course would be to assume that
the grains always drift at their equilibrium drift velocity and to
perform a ``1.5 fluid'' calculation. It turns out, however, to be
difficult to determine whether or not the assumption of equilibrium
drift is justified, c.f. Section \ref{sec:stationarymodels}.  For a
discussion about the comparison of two--fluid and ``1.5 fluid''
calculations see Appendix \ref{sec:append}.
%
%----------------------------------------------------------------------------
%
\section{Numerical calculations}
\label{sec:results}
\subsection{Numerical method}
\label{sec:numericalmethod}
The continuity and momentum equations are solved using an explicit
scheme. A hydrodynamics code was specially written for this purpose.
It uses centered dif\-fer\-en\-cing and a two--step,
predictor--corrector scheme, applying Flux Corrected Transport (FCT)
(\cite{Boris:1976}). Second order accuracy is achieved for the single
fluid and momentum coupled (``1.5 fluid'') calculations. In the two
fluid computation we applied, whenever needed, Local Curvature
Diminishing (LCD) (\cite{Icke:1991}), at the risk of introducing first
order behavior.
\subsection{Initial and boundary conditions, grid}
\label{sec:initinner}
\begin{figure*}
	\resizebox{\hsize}{!}{\includegraphics{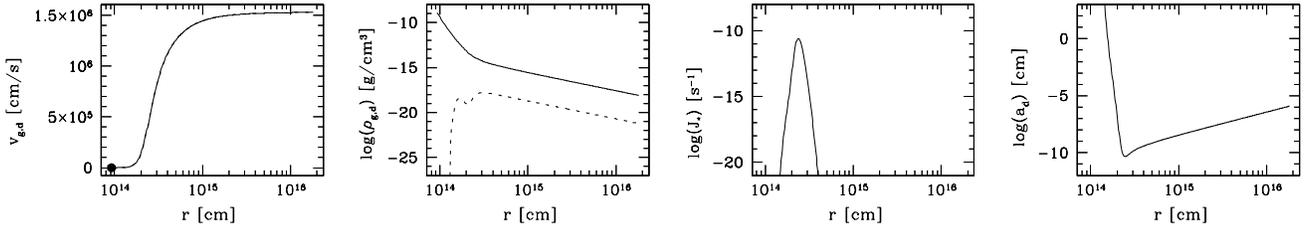}}
	\caption{Velocity (no drift), gas and dust density, nucleation
	rate and average grain radius for the initial profile.}
	\label{fig:initmodel}
\end{figure*}

As an initial model for the calculation, a stationary profile for
\object{IRC +10216}, kindly provided by J.M. Winters
(\cite{Wintersea:1994}), was used, see Fig.~\ref{fig:initmodel}.
%
%-----------------------------------------------fig:initmodel=simis10276_f1
%--------------------------------------------------------------------------
%
Stellar parameters of this model are: $M_{*} = 0.7 M_{\odot}$, $L_{*}
= 2.4 \cdot 10^4 L_{\odot}$, $T_{*} = 2010 \mbox{K}$ and a carbon to
oxygen ratio $\epsilon_\mathrm{C}/\epsilon_\mathrm{O} = 1.40$. The
corresponding stellar radius is $R_{*} = 9.20 \cdot 10^{13}
\mbox{cm}$, $R_\mathrm{max} = 200 R_{*}$. The mass loss rate for the
initial model is $\dot{M} = 8 \cdot 10^{-5} M_{\odot}
\mbox{yr}^{-1}$. In order to compare our calculations with
observations, we extend the computational grid to 1287
$R_{*}$. Because no initial data is known for the grid extension, we
simply set the initial values for $r > 200 R_{*}$ of all flow
variables equal to their value at $r = 200 R_{*}$. As a consequence of
this, a transient solution will have to move out of the grid before
the physically correct solution can settle.\\
Grid cells are not equally spaced, since a high resolution is
desirable in the subsonic area but not necessary in the outer
envelope. The grid cells are distributed according to:
\begin{equation}
\frac{r[n] - r[n-1]}{r[1] - r[0]} = q^{n-1/n_\mathrm{max}-1}
\end{equation}
The number of cells in the grid, $n_\mathrm{max}$, used here is 737
and the size ratio $q$ between the innermost and the outermost cell is
318.\\
One of the most important aspects of a numerical hydrodynamics
calculation is the treatment of the inner boundary. Since the (long
time averaged) mass fluxes throug the inner and the outer boundary
must be equal, setting the inner boundary essentially means fixing the
mass loss rate. We have, in our calculations, fixed the density and
velocity in the innermost grid cells, so that the {\em advective} mass
and momentum fluxes (i.e. the first order derivatives of the flow
variables) through the inner boundary are constant. Note that the
temperature was constant as a function of time as well so that also
the pressure will be fixed. In reality, however, velocity and density
will vary with time. To account for a variable inflow of mass into the
envelope, we permit also {\em diffusive} inflow of mass. This flow
depends upon second order derivatives near the inner boundary and
therefore models quite realistically the cause of matter inflow into
the envelope.  At the inner boundary, the main driving term of the
wind is not yet active and the velocities are very small because newly
formed small grains, which are very sensitive to radiation pressure,
are formed farther out. Therefore, the oscillations of the envelope
are clearly not caused by the implementation of the inner boundary.\\
To model the diffusive flux at the inner boundary, we could have
introduced a separate diffusion term. There is no need to do so,
however, since our numerical scheme involves the calculation of a
diffusion term already. This diffusion term ({\em numerical
viscosity}) is part of our finite difference scheme and it is locally
(i.e. at extrema) required to stabilize the centered differencing
method. Whenever numerical viscosity is not strictly needed to
stabilize the numerical scheme it will be canceled by an
anti-diffusion term (\cite{Boris:1976}). A detailed description of
this method is beyond the scope of this paper, for details the reader
is referred to \cite*{Icke:1991}. We want to allow for diffusion at
the inner boundary. Instead of adding explicitly a diffusion term we
can simply somewhat reduce the anti--diffusion at the inner
boundary. That way, not all of the numerical diffusion is canceled and
effectively a diffusive flux is created at the inner boundary.\\
Although important for the AGB evolution, no stellar pulsations or
time dependent luminosities were used. Often, in hydrodynamical
simulations of late type stars, stellar pulsations are introduced as a
time dependent inner boundary condition. In the absence of pulsations,
the average grain near the inner boundary will be large.  Since larger
grains are less efficiently accelerated by the radiative force than
smaller ones, the stationary inner boundary condition will lead to
small velocities in the lower envelope. As a result of the inefficient
radiative force on large grains, these grains will also tend to drift
at high or even non--equilibrium drift speeds. To avoid this unwanted
behavior, equilibrium drift is imposed in the first 2.8 $R_{*}$, also
in the two--fluid calculation.
\subsection{Calculations}
In order to determine the effect of grain drift on the outflow, we
perform three types of calculation. First, we solve the full
two--fluid system including gas chemistry, grain formation and growth
and the continuity and momentum equation for both gas and grains. The
viscous momentum transfer during each numerical time step is
calculated by integration of $f_\mathrm{drag}$ over this time step as
was presented in Section \ref{sec:momcoup}. Division by the duration
of the time step gives an expression for $f_\mathrm{drag}$ that can be
inserted in the momentum equations, Eqs.(\ref{eq:momg},\ref{eq:momd}). When
solving, the left hand side of these equations is multiplied by the
time step again, so that indeed the correct amount of momentum is
transferred.\\
Next, a 1.5--fluid calculation is performed. Here, the drag force is
calculated by assuming equilibrium drift in
Eq.(\ref{eq:fdragsimple}). The dust velocity is taken to be the sum of
the gas velocity and equilibrium drift velocity, according to
Eq.(\ref{eq:vdrifteq}). The momentum equation of the dust is not
solved.\\ 
Finally, we also perform a single fluid calculation. Here too, only
the gas momentum equation is solved. The drag force exerted on the gas
is taken to be equal to the radiation force on the grains. Now, the
velocity of the grains is simply set equal to the gas velocity. From
the 1.5 and single fluid calculations, we expect to learn about the
influence of (non--equilibrium) drift on the flow, when comparing them
to the two fluid calculation.\\
All three models were evolved $10^6$ numerical time steps, which amounts to 
$9.71 \cdot 10^{10}$, $1.67 \cdot 10^{11}$ or $3.14 \cdot 10^{11}$
seconds, depending on the model.
\subsection{Results}
\label{sec:resultsubsection}
\begin{figure*}
	\resizebox{\hsize}{!}{\includegraphics{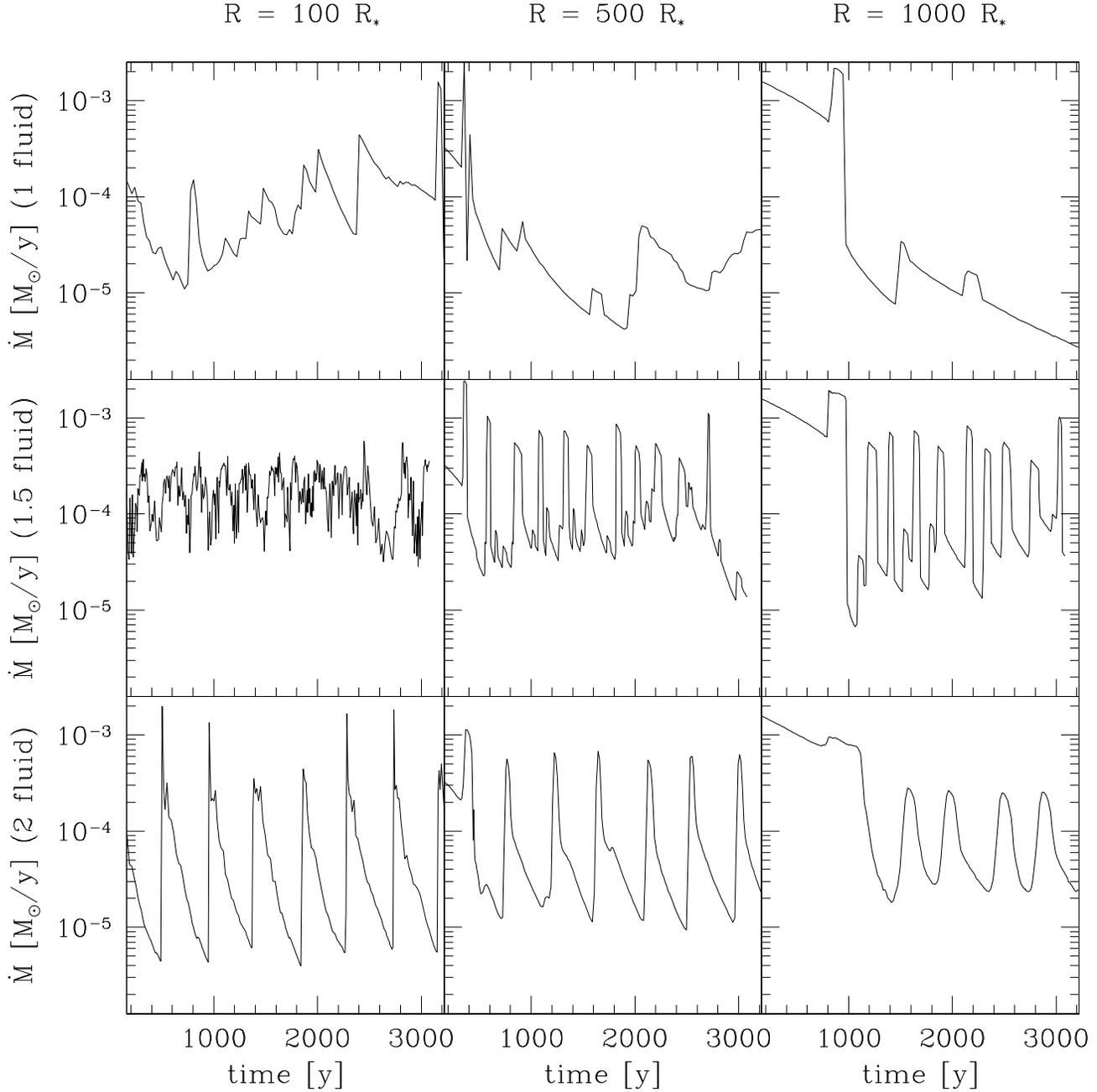}}
	\caption{From top to bottom: Mass loss rates for single fluid
	(no drift, gas and grain have equal velocity, ``position
	coupling''), 1.5--fluid (equilibrium drift, gas and grains
	have equal acceleration, ``momentum coupling'') and two--fluid
	(no assumptions on drift, no coupling imposed) calculations,
	for R = 100, 500 and 1000 $R_\mathrm{*}$. Note that the first 150
	years of output in the 500 $R_\mathrm{*}$ plot and the first 800
	years in the 1000 $R_\mathrm{*}$ plot show the passing of the
	transient solution due to the extension for the calculational
	grid w.r.t. the intial model.} \label{fig:mdot_gd}
\end{figure*}

Fig.~\ref{fig:mdot_gd} shows the mass loss rate at $R$ = 100, 500 and
1000 $R_\mathrm{*}$ as a function of time for the three
calculations. The first 150 years of output in the 500 $R_\mathrm{*}$
plot and the first 800 years in the 1000 $R_\mathrm{*}$ plot show the
passing of the transient solution. This is a result of extending the
grid from 200 $R_\mathrm{*}$ in the initial profile to 1287
$R_\mathrm{*}$ in the calculation, the flow needs some time to reach
the additional gridpoints.\\
Both the 1.5 and the two--fluid model show quasi--periodic
oscillations. From plots which cover a longer time interval (not shown
here) we infer that the variations in the mass loss rate in the single
fluid calculation behave quasi--periodically as well, on a time scale
of a few thousand years. An immediate conclusion from this is, that
the presence of grain drift is important for variations of the mass
loss rate.\\
The time between two peaks in the mass loss is approximately 200 to
350 years for the 1.5--fluid model, and about 400 years for the
two--fluid model. Both numbers lie nicely in the range of the
separation of 200--800 years between the shells that
\cite*{MauronHuggins:1999} observed in \object{IRC +10216}. \\
In all three calculations we see that the short time variations that
are present at 100 $R_\mathrm{*}$, have disappeared far away from the
star. \cite*{MauronHuggins:2000} note that this ``wide range of shell
spacing, corresponding to time scales as short as 40 yr (close to the
star) and as long as 800 yr'', should be accounted for in a consistent
model. This poses no problems, since the disappearance of the smaller
scale structures is simply due to dispersion and hence will appear in
any flow in which perturbations do not propagate with exactly the same
speed.\\
The fact that the two--fluid calculation shows less variations on
short times scales than the 1.5--fluid model may be due to the more
first order character of the former (as a result of the LCD term, see
Section \ref{sec:numericalmethod}). We shall see that in the
two--fluid calculation, in large parts of the envelope, grains move at
their equilibrium drift velocity.
The time averaged mass loss rate, estimated from
Fig.~\ref{fig:mdot_gd}, lies around $\dot{M} = 1 \cdot 10^{-4}
M_{\odot} \mbox{yr}^{-1}$. The fact that this is somewhat higher than
the mass loss rate of the initial model indicates that indeed the
diffusive flux at the inner boundary has contributed, see Section
\ref{sec:initinner}. Our limited implementation of the radiative force
(we use a grey dust opacity and take the extinction cross section of
the grains equal to the geometrical cross section) causes the
velocities in our calculation to be higher than the velocities in the
initial model. Using a lower value for the stellar luminosity
(e.g. using the core mass--luminosity relation) has proven to
immediately lower the outflow velocity and hence the mass loss rate.\\
%
%-------------------------------------------------fig:mdot_gd=simis10276_f2
%--------------------------------------------------------------------------
%
%----------------------------------------fig:greycoplotsvel=simis10276_f3
\begin{figure*}
	\resizebox{\hsize}{!}{\includegraphics{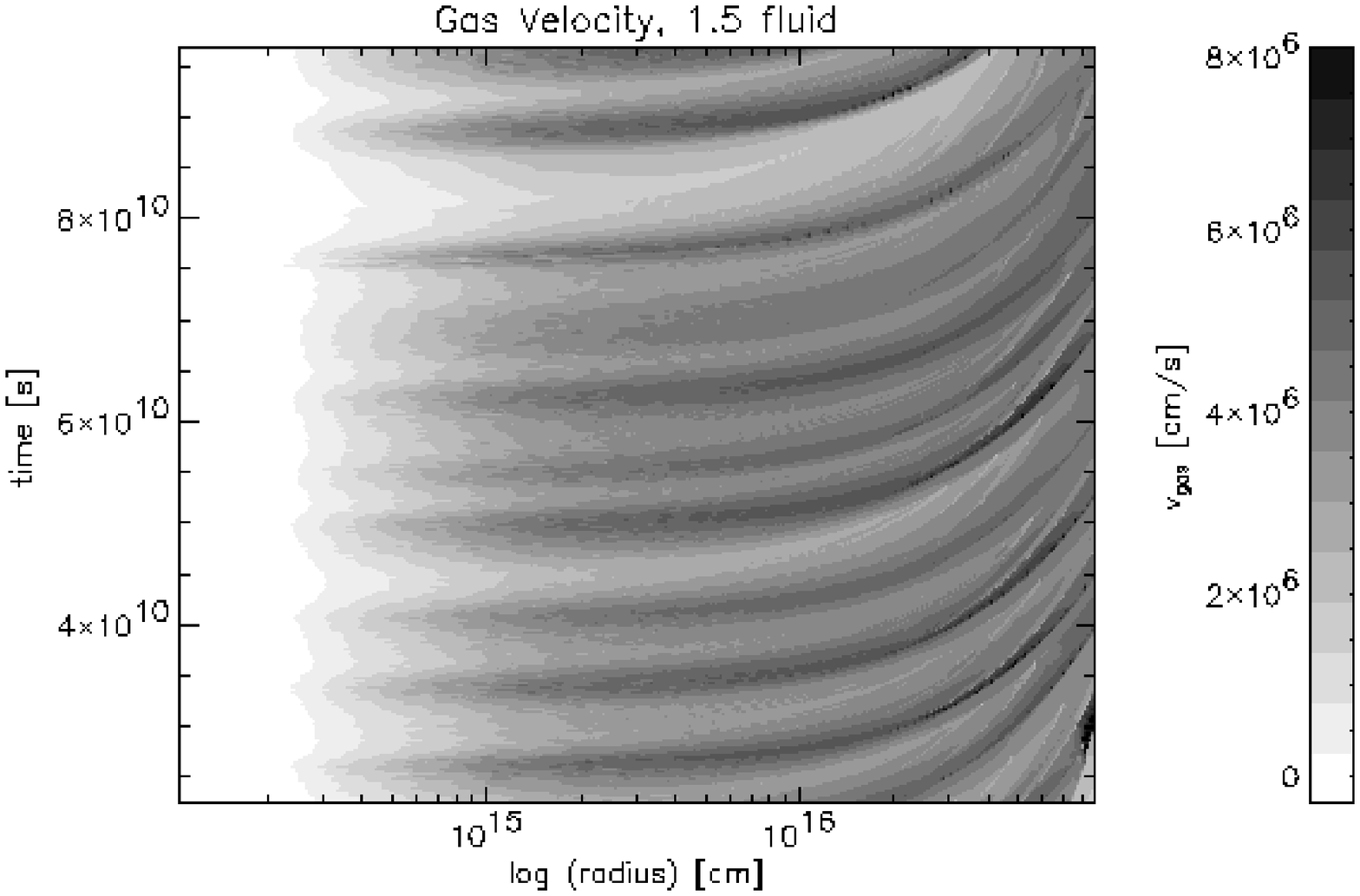}\includegraphics{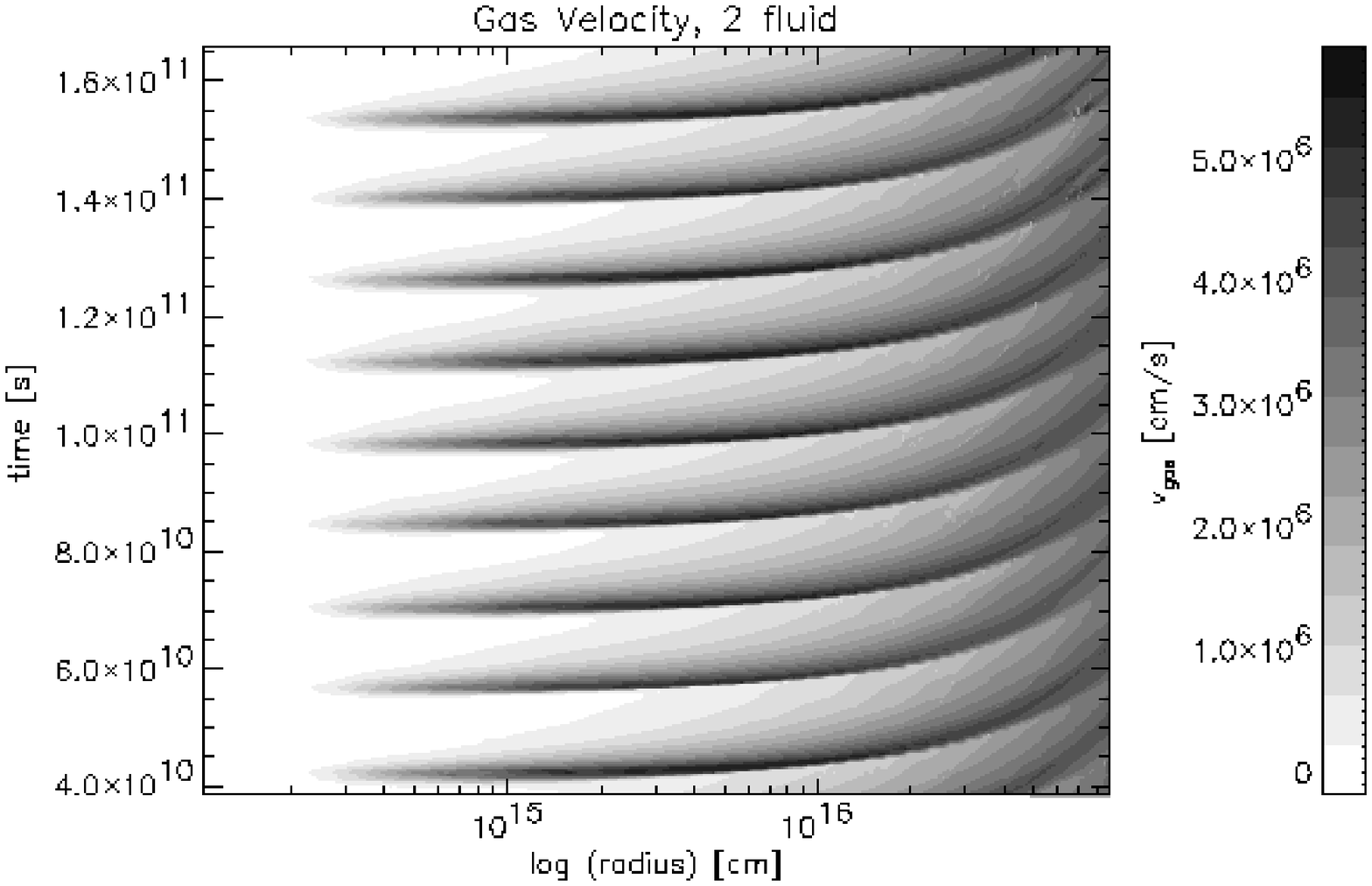}}
	\resizebox{\hsize}{!}{\includegraphics{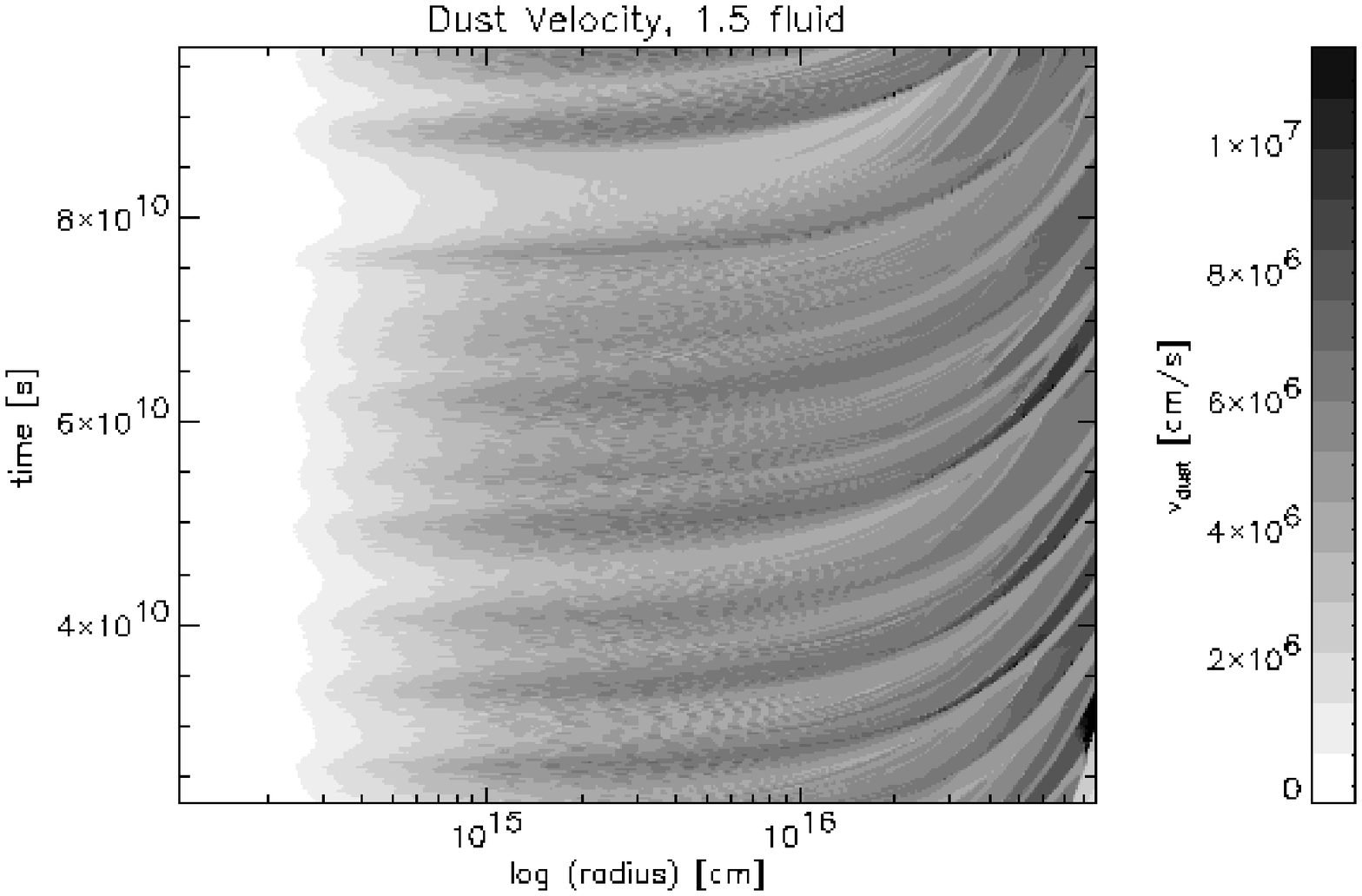}\includegraphics{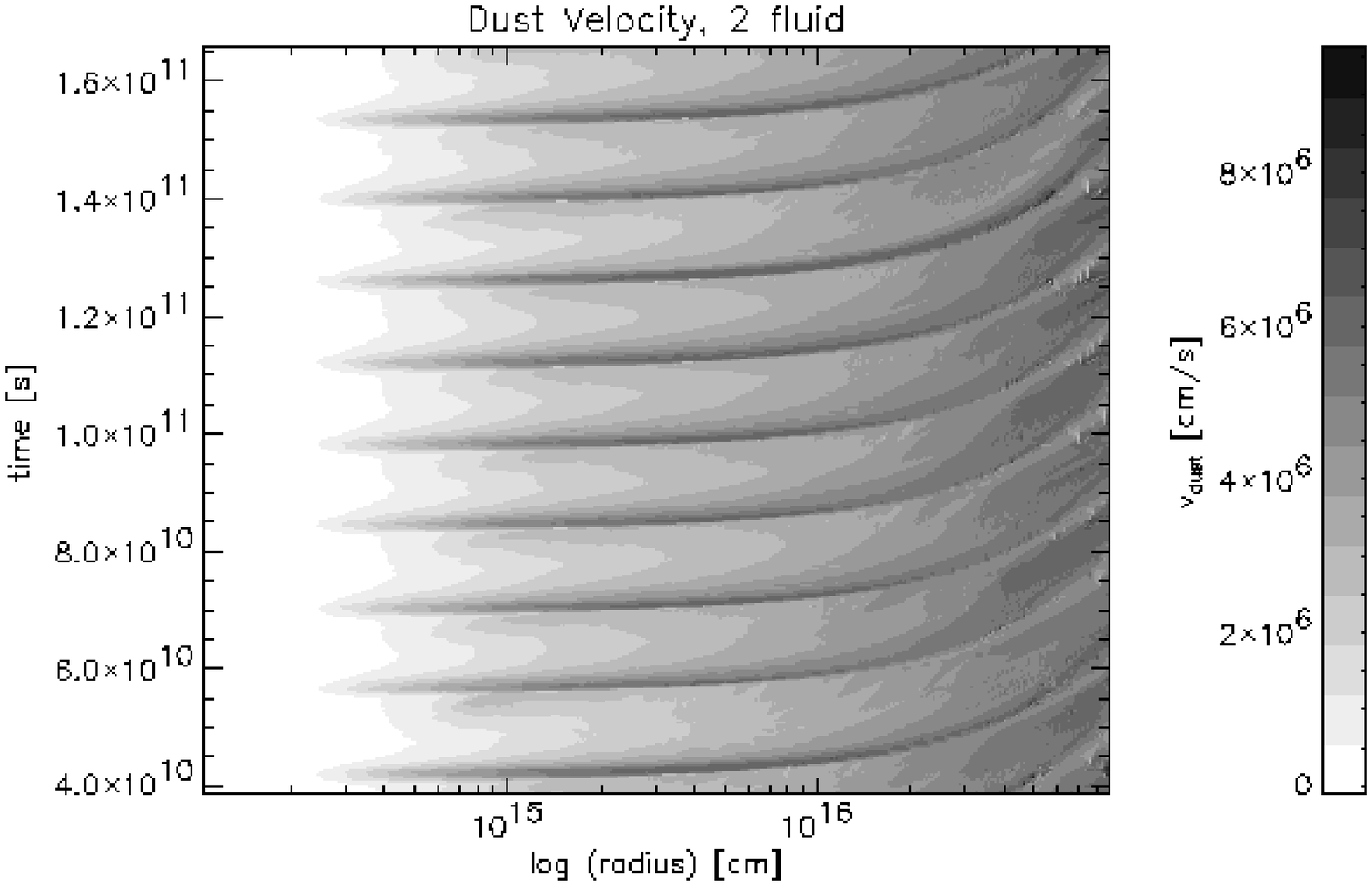}}
	\caption{Gas and dust velocities as a function of radius and time for the 1.5 and the two--fluid model.}
         \label{fig:greycoplotsvel}
\end{figure*}
%--------------------------------------------------------------------------
%
%----------------------------------------fig:greycoplotsvel=simis10276_f4
\begin{figure*}
	\resizebox{\hsize}{!}{\includegraphics{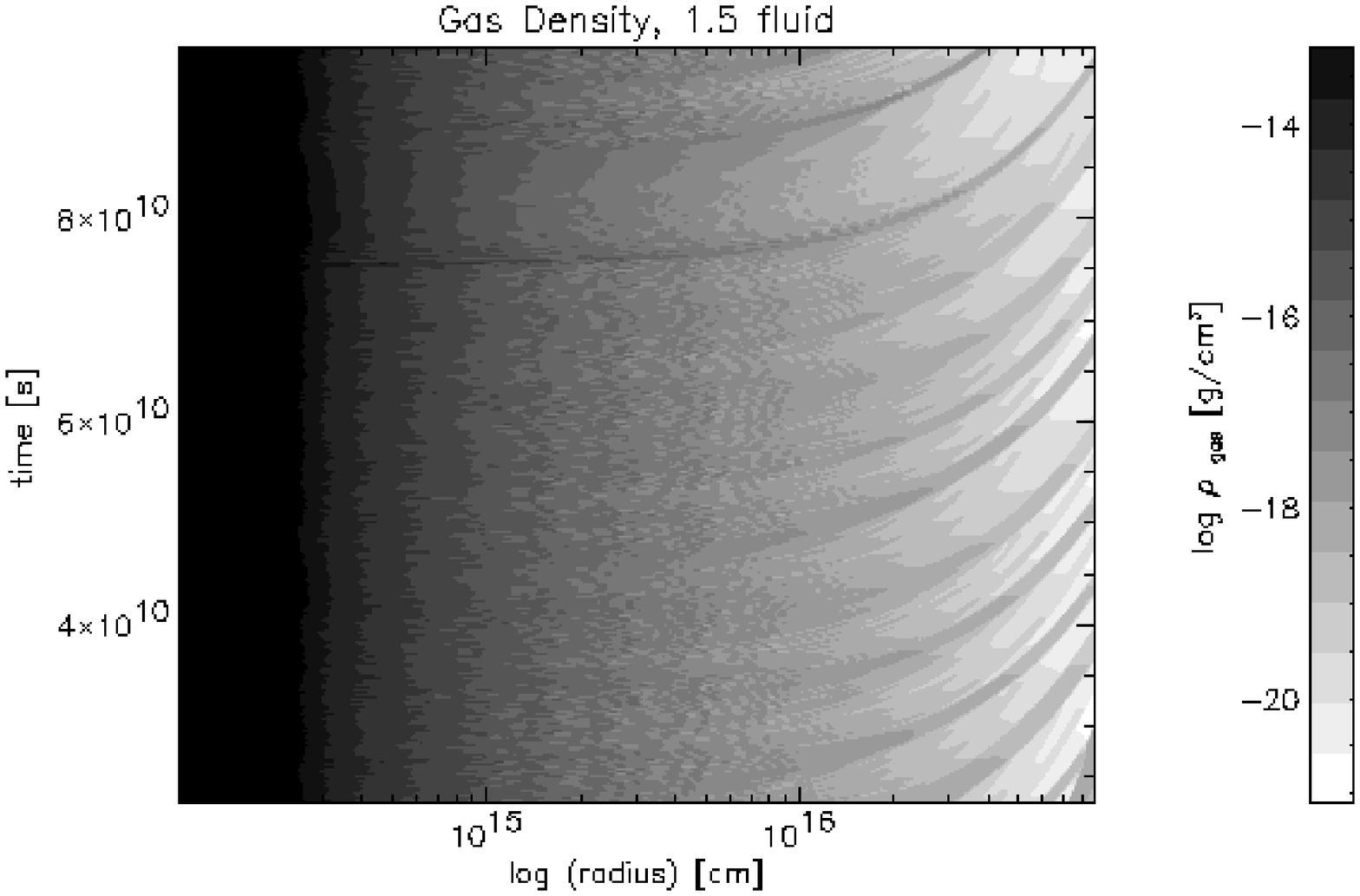}\includegraphics{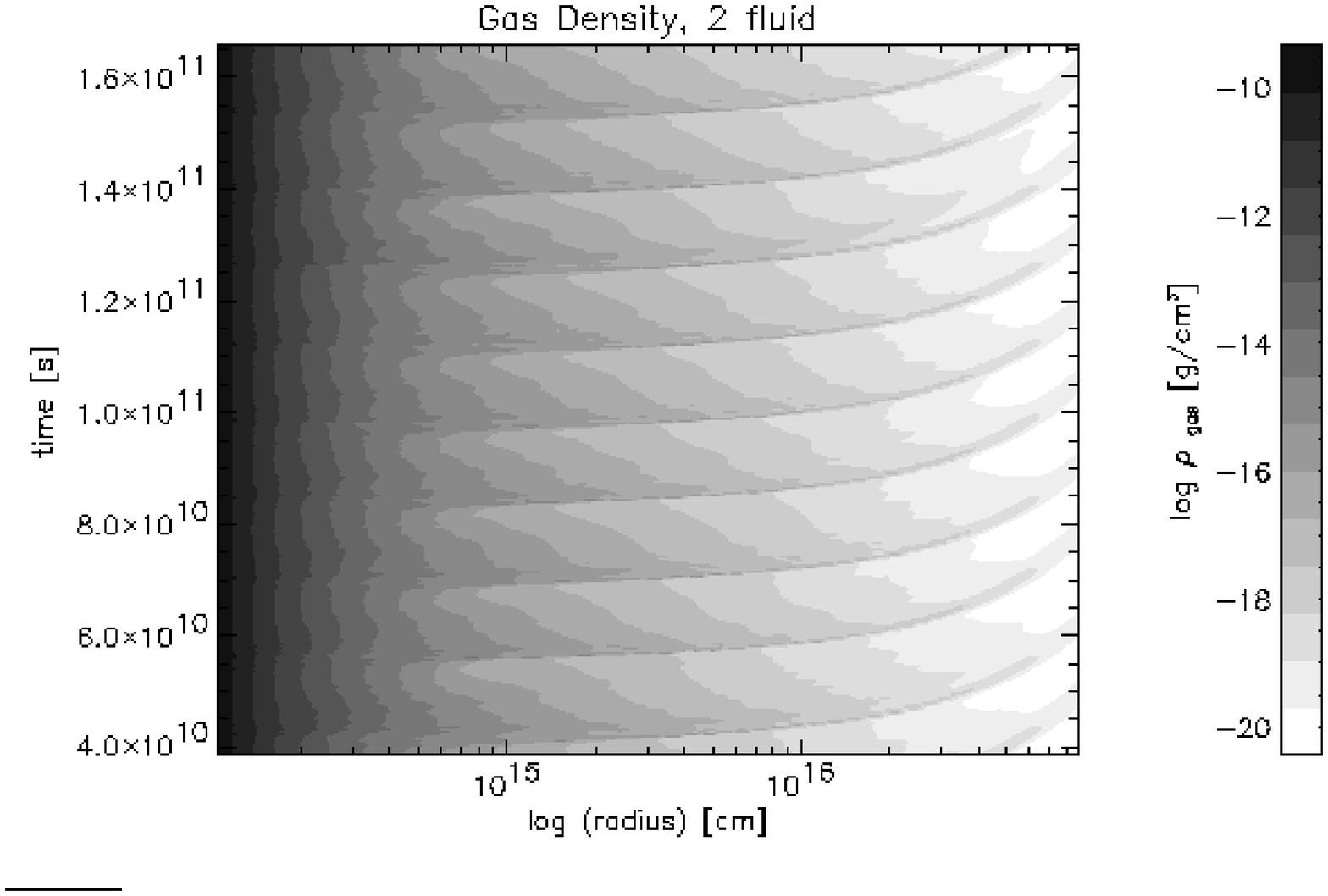}}
	\resizebox{\hsize}{!}{\includegraphics{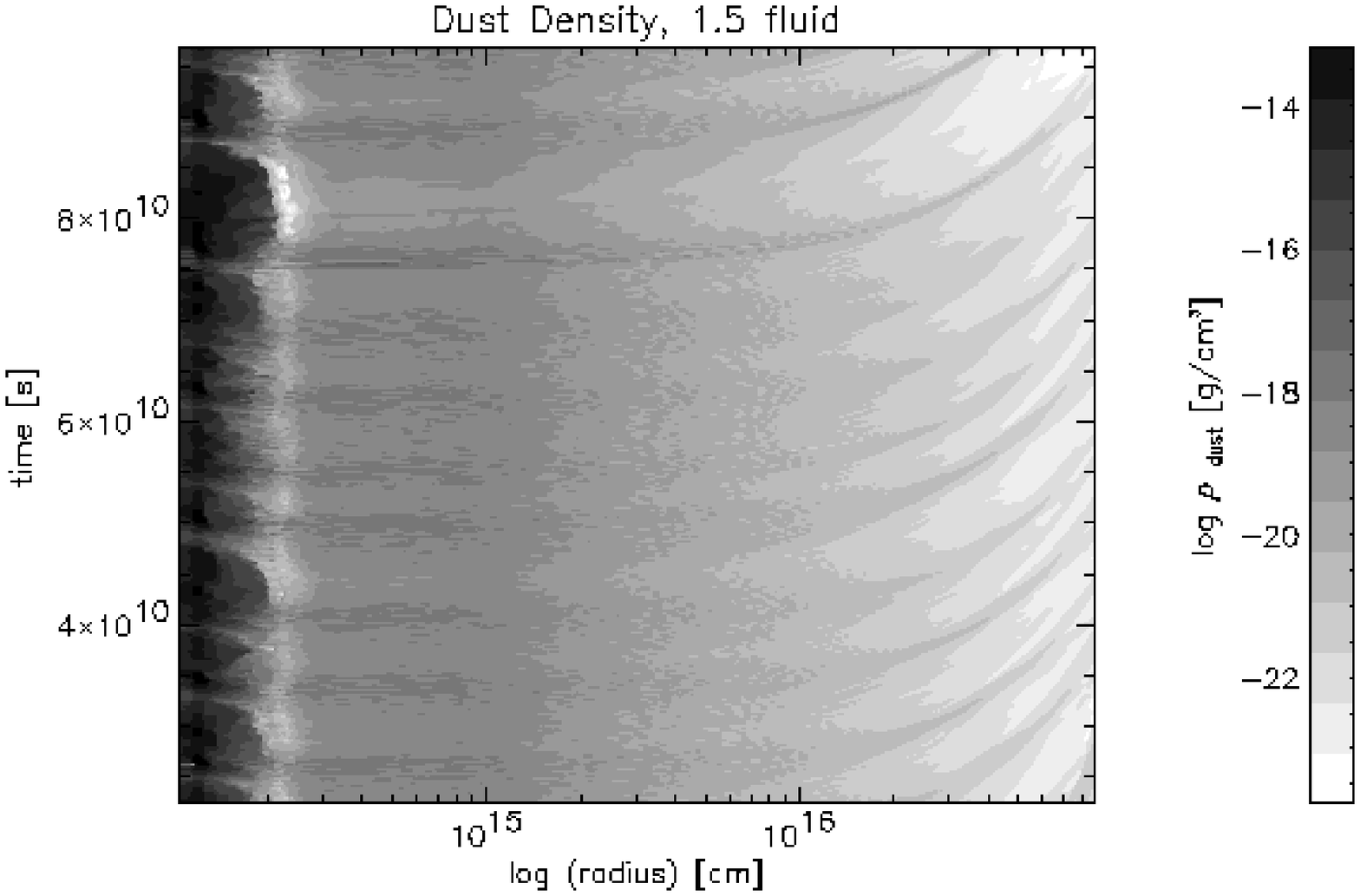}\includegraphics{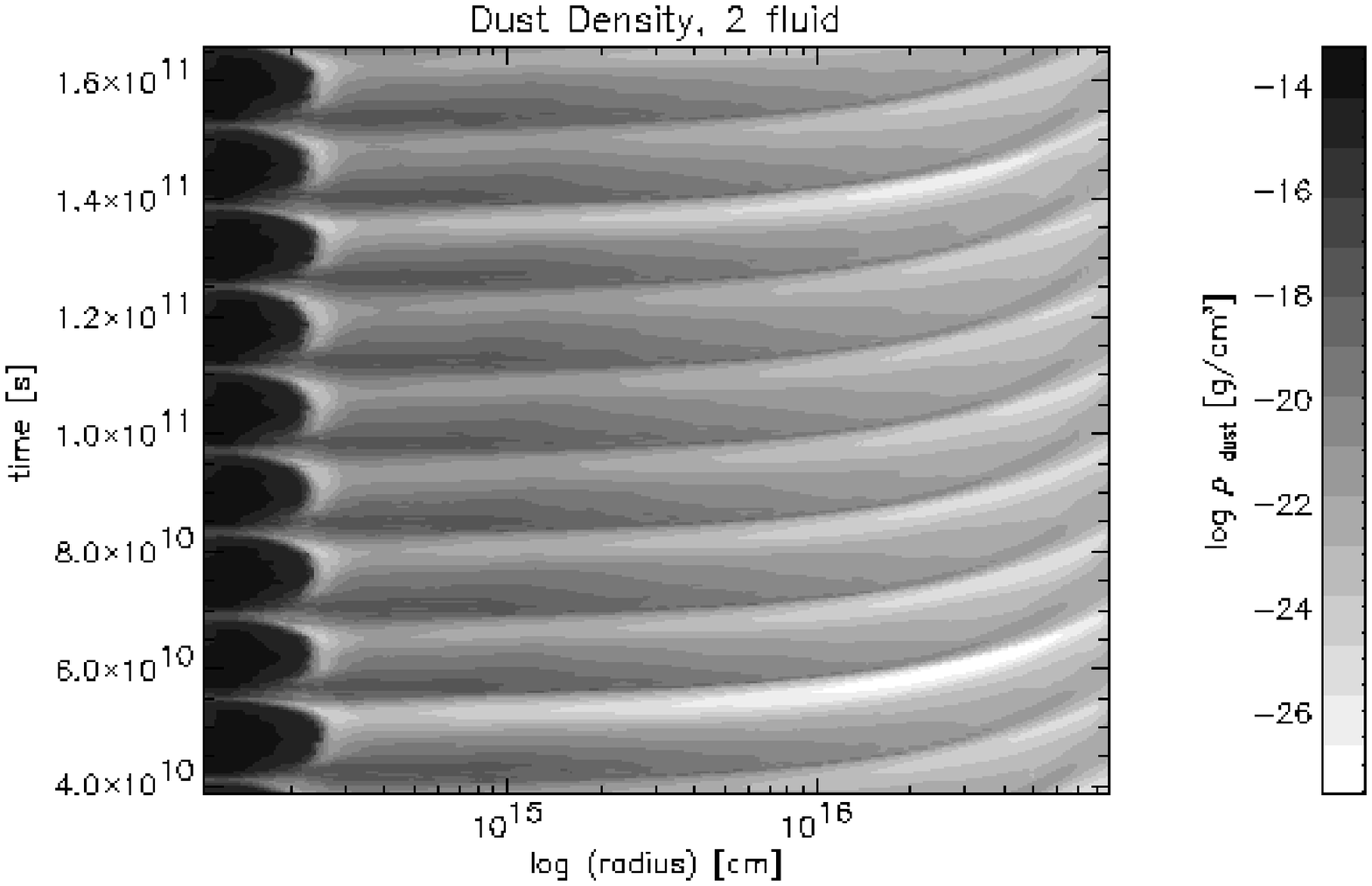}}
	\caption{Gas and dust densities as a function of radius 
and time for the 1.5 and the two--fluid model.}
         \label{fig:greycoplotsdens}
\end{figure*}
%--------------------------------------------------------------------------
%
\noindent
Figs.~\ref{fig:greycoplotsvel} and \ref{fig:greycoplotsdens} show, for
the 1.5 and the two--fluid model, the gas and dust velocities and
densities, as a function of radius and time. Throughout the whole
grid, the fluctuations occurring in the two--fluid calculation are
more regular that those in the 1.5--fluid model.  The velocities of
gas and dust in the momentum coupled calculation reach values that are
up to 25\% higher than in the two--fluid calculation. In the latter,
matter is less accelerated than in the former, especially for radii
larger than about $2 \cdot 10^{16}$ cm. Probably, this is a result of
non--equilibrium drift, which starts to appear around this radius (see
Fig.~\ref{fig:cycle2f}).  Non--equilibrium drift occurs when the time
needed by a grain to reach its equilibrium drift velocity is long
compared to the dynamical time scale.  During a period of
non--equilibrium drift, the gas is not being maximally accelerated and
both gas and dust velocities will be lower than in a phase of
equilibrium drift.\\
The gas density structure (Fig.~\ref{fig:greycoplotsdens}) for the
1.5--fluid and the two--fluid calculation look similar. The main
difference is that short time scale variations are present in the
lower regions of the former, whereas large scale effects dominate the
latter. The density structure plots for the dust show another
difference: the perturbations in the 1.5--fluid flow appear as local
increments of the density but in the two component flow the variations
rather look like dips in the average profile. Maximum outflow density
for gas and grains are in phase in the two--fluid model though, the
``dust pulse'' is significantly broader than but centered around the
maximum in the gas outflow. This is not just the case in the upper
parts of the envelope, where non--equilibrium drift is present, but
also for smaller radii.\\
%
%------------------------------------------fig:cycle15f=simis10276_f5
\begin{figure*}
	\resizebox{\hsize}{!}{\includegraphics{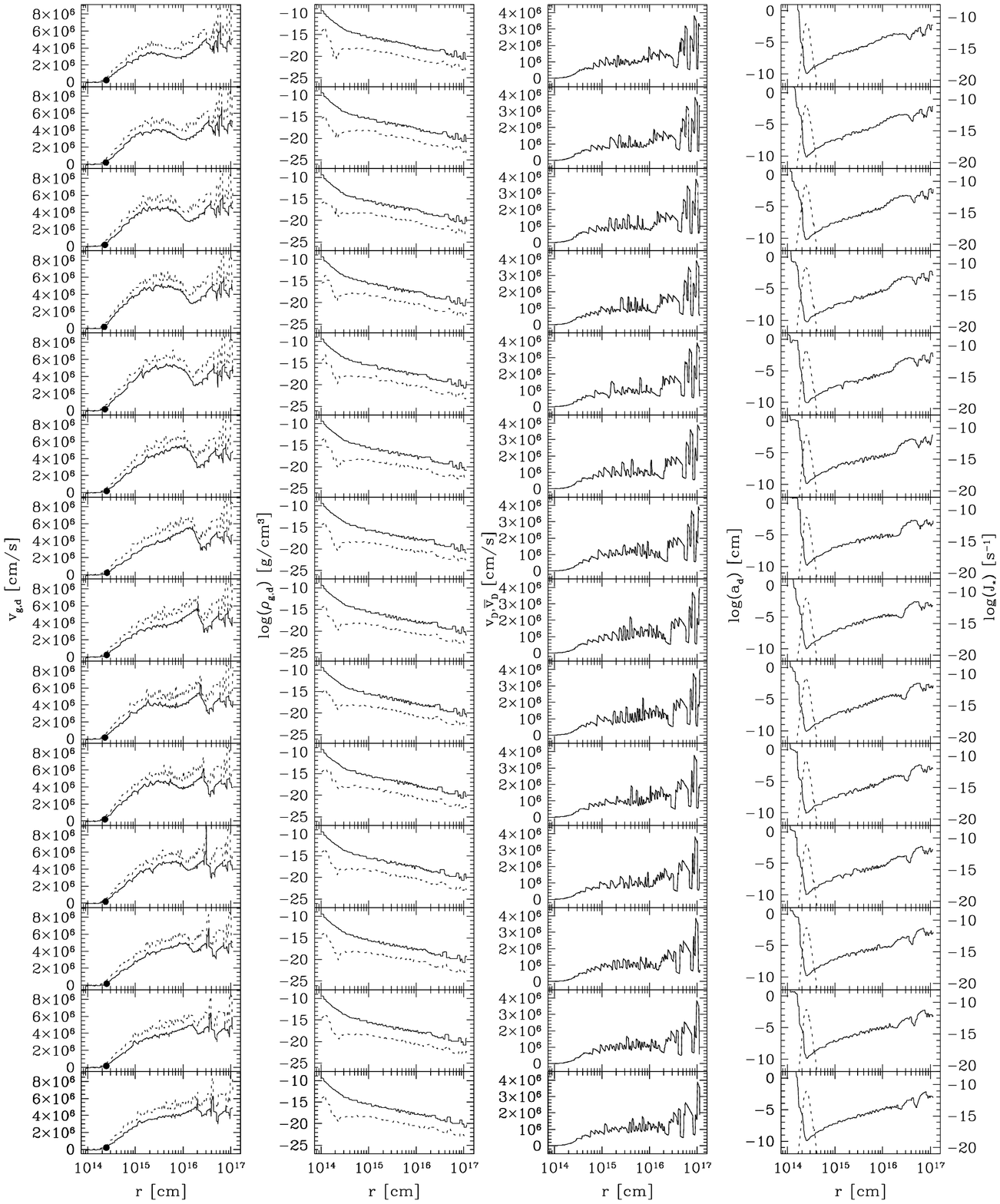}}
	\caption{1.5--fluid model. First column: gas and dust velocity
	(dashed line).  The dot denotes the location of the critical
	point. Second column: gas and dust density (dashed line).
	Third column: drift velocity. Fourth
	column: average grain radius and grain nucleation rate (dashed
	line). The frames show (from top to bottom) the flow profile
	at 0, 30, 56, 81, 105, 132, 164, 197, 225, 252, 280, 310,
	352 and 404 years after the first frame.}
	\label{fig:cycle15f}
\end{figure*}
%----------------------------------------------------------------------
%
%------------------------------------------- fig:cycle2f=simis10276_f6
\begin{figure*}
	\resizebox{\hsize}{!}{\includegraphics{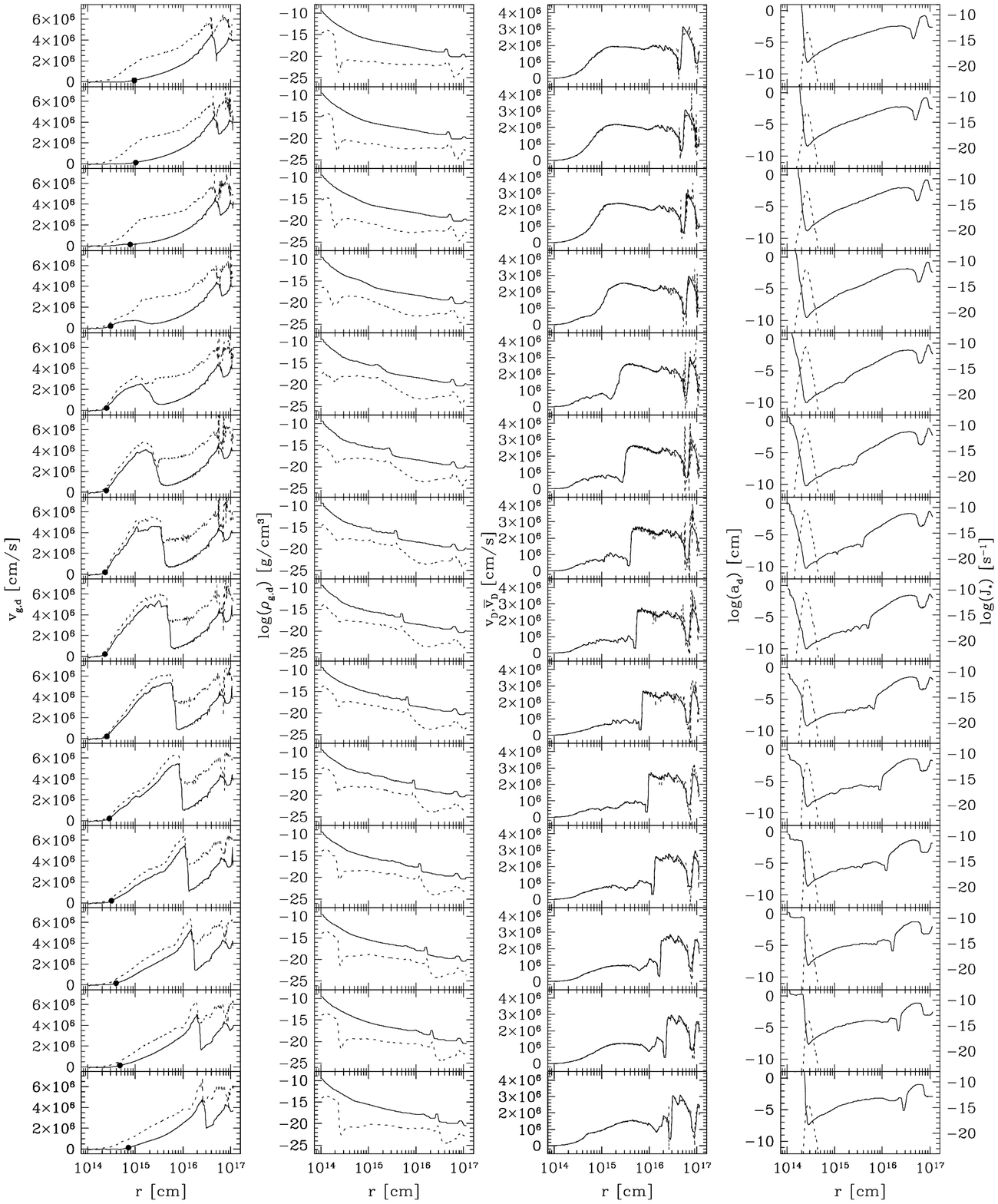}}
	\caption{Two--fluid model. First column: gas and dust velocity
	(dashed line). The dot denotes the location of the critical
	point. Second column: gas and dust density (dashed
	line). Third column: drift velocity (dashed line) and
	equilibrium drift velocity (full line).  Fourth column:
	average grain radius and grain nucleation rate (dashed
	line). The frames show (from top to bottom) the flow profile
	at 0, 32, 63, 93, 116, 131, 144, 156, 170, 189, 214, 245,
	284 and 330 years after the first frame.}
	\label{fig:cycle2f}
\end{figure*}
%--------------------------------------------------------------------------
%
In Figs.~\ref{fig:cycle15f} and \ref{fig:cycle2f} we plot a series of
snapshots, displaying the evolution of various flow variables during
one instability cycle for the 1.5 and the two--fluid model.  For the
1.5--fluid calculation the drift velocity is, by definition, always
equal to its equilibrium value, which shows a time dependent
behavior. In the two--fluid flow we find that the drift velocity, out
to approximately $10^{16}$ cm, equals the equilibrium value. At larger
radii, small deviations from equilibrium drift are detected.\\
We want to stress that the fact that we see equilibrium drift in the
lower and intermediate regions of the two component model {\em only}
implies that equilibrium drift is established on a time scale shorter
than the dynamical time scale. It does {\em not} however exclude the
possibility that non--equilibrium drift occurs on shorter time scales,
see Appendix \ref{sec:append}.
\subsection{The origin of the mass loss variability}
\label{sec:pulsationorigin}
To investigate what causes the variability we will step through the
frames of Fig.~\ref{fig:cycle2f} for the two--fluid
calculation. Thereafter, we will discuss the differences with the 1.5
fluid model. The mass loss rate of a stellar wind is determined in the
subsonic region (see e.g. \cite{LamersCassinelli:1999}), therefore in
the following, when investigating the mechanism underlying the
variability, we focus on this region, unless explicitly mentioned.\\
In Fig.~\ref{fig:cycle2f}, first frame, we see that the onset of the
mass loss variability is the situation in which the dust has a
velocity that is significantly higher than the gas velocity. This
means that the residence time of a grain in the parts of the envelope
where grains can grow is relatively short so that the average grain
size will be on the small side. The smaller the grain, the more
efficient radiation pressure will be, since small grains have a large
surface to mass ratio and since we have assumed that the grain
extinction cross section equals the geometrical cross section.  Hence,
radiative acceleration of grains is efficient and the velocity of the
small grains increases further. Because position coupling is not
imposed, the gas velocity can stay low and the drift velocity
increases. Meanwhile (frames 2 and 3), the average grain radius
decreases, grain acceleration becomes more efficient, the dust
velocity grows, grains become smaller, and so forth. Also, the total
mass density of the dust component in the innermost region decreases.
When the grain radius in the subsonic region drops below a certain
critical value, momentum transfer from grains to gas becomes efficient
and the gas is accelerated (frame 4).  This results in an increase of
the gas density and hence of the number density of condensible
particles. Since the grain nucleation rate is extremely sensitive to
the molecular abundances, this results in an immediate increment of
the nucleation rate (frame 4). The new production of condensation
kernels leads to a further decrease of the average grain radius and an
increase of the total grain mass density. Due to the large abundance
of small grains, radiative acceleration and the transfer of momentum
from grains to gas are very efficient, so that both gas and grains
move out with high velocities (frames 5--8). On their way out, the
small grains concentrate in a narrowing shell, since the decrease of
the average grain radius in time coincides with an increase of their
velocity. The gas develops a shell at the same time, as a result of
the forming shock.  The normal, Parker--type, stellar wind profile is
now visible. We will refer to this phase as the ``fast phase'' (frame
5--9).  Though not very clear from the figure, at the same time, a
rarefaction wave moves in the opposite direction, leading to a
decrease of the gas density, and of the number densities of the
condensible species, below the sonic point. Although the density
decrease is not so big, the nucleation rate reacts instantaneously
(frames 9--13), showing a strong decrease traveling from the sonic
point inwards. Hence, the passing of the rarefaction wave is
immediately visible in the increase of the average grain radius
because the production rate of new small grains decreases (frames
9--13).  This illustrates the enormous sensitivity of the nucleation
rate on the densities. The gradual increase, in time, of the average
grain radius, brings about a less efficient radiative acceleration of
the dust, hence a decrease of the grain velocity and a further
increase of the grain radius, and so forth. This we will call the
``slow phase'' of the variability cycle (frames 10--14 and 1--4).  Due
to the larger grain size, the momentum transfer between grains and gas
becomes less efficient, resulting in larger drift and dust velocities
(frame 14).  This brings us back to the situation in the first
frame.\\
Crucial in the process of shell formation as described above are the
two ``turn--around'' points, at which the nucleation rate starts to
increase and decrease. First, at the end of the fast phase, the
passage of the rarefaction wave triggers the end of a period of high
nucleation rate. In the slow phase the gas--grain coupling has becomes
less efficient, due to the larger average grain size.  Grains then
reach a higher drift velocity, become smaller and will again transfer
their momentum efficiently to the gas, so that the latter can
accelerate, increasing the density. This gives rise to favourable
circumstances for grain nucleation again. Clearly, the behavior of the
system during the slow phase is dominated by the existence of grain
drift.  This immediately explains why variability in the mass loss
rate in a single fluid system is less well regulated (see 
Fig.~\ref{fig:mdot_gd}).\\
When comparing Fig.~\ref{fig:cycle2f} and Fig.~\ref{fig:cycle15f}, the
absence of the slow phase in the variability cycle in the latter
strikes the eye. This can be attributed to the imposed equilibrium
drift in the 1.5--fluid flow. In the two--fluid system the drift
velocity is directly influenced by the dynamics. In the 1.5--fluid
model, however, the (equilibrium) drift velocity is only indirectly
determined by the dynamics, namely via the (number) densities and the
grain size. The fact that the variability character is still observed
in this calculation is a consequence of the fact that the drift
velocity, although not actively, does change as a function of time, in
combination with the extreme sensitivity of the nucleation rate to the
density and of the dynamics, via the drag force, on the grain size,
and density.  The sensitivity of the system is well visible in
Fig.~\ref{fig:cycle15f}: any variation of the densities, grain size
and nucleation rate is hardly visible (also because they are plotted
logarithmically, ranging over many orders of magnitude) but the
resulting variations in the velocity field are clearly present.
\subsection{Comparison with observations}
\label{sec:compareobs}
%
%----------------------------------- fig:greycoplotsvel=simis10276_f7
\begin{figure*}
\centering
	\includegraphics[width=4.25cm]{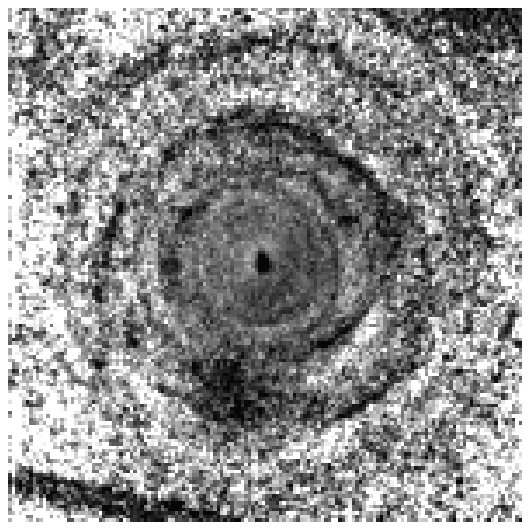}
	\includegraphics[width=4.25cm]{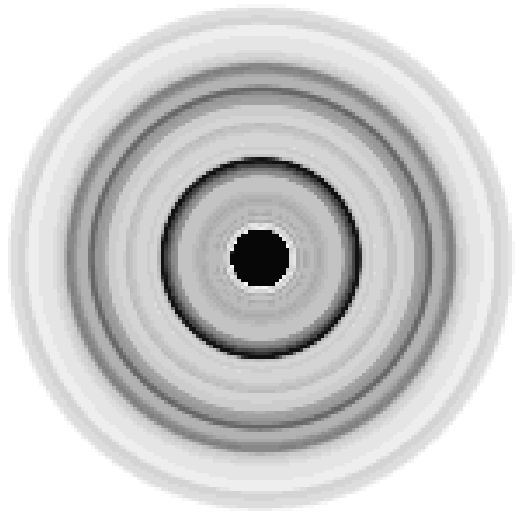}
	\includegraphics[width=4.25cm]{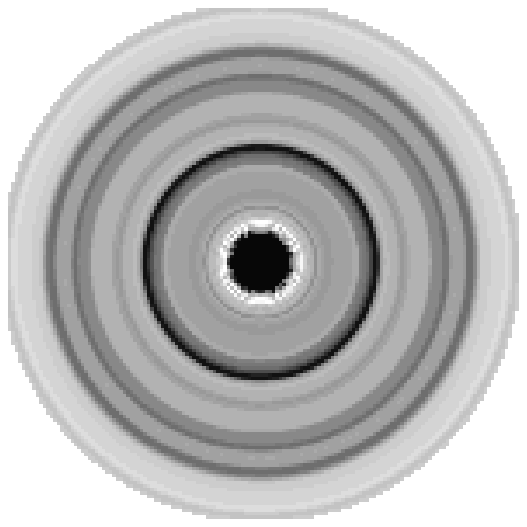}
	\includegraphics[width=4.25cm]{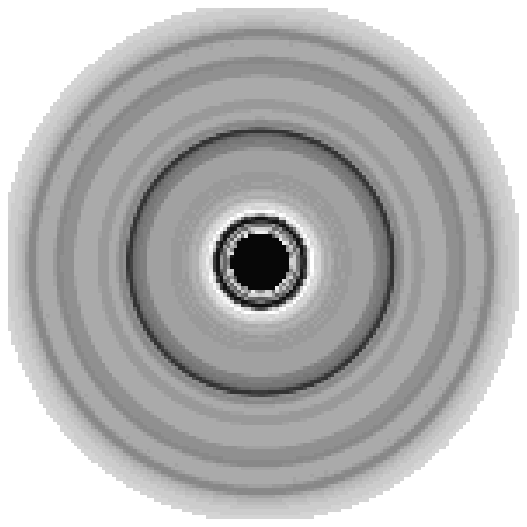}
	\caption{Upper left frame: Composite $B + V$ image of
	IRC +10216, with an average radial profile subtracted
	to enhance the contrast (adapted from
	Mauron \& Huggins (1999)). Note that a few patches in the
	image are residuals of the removal of the brightest
	background objects, and these should be ignored. Other frames:
	series of snapshots of our 1.5--fluid calculation. Plotted is
	the dust column density, also with an average radial profile
	subtracted.  The average radial profiles are calculated for
	each snaphot separately, hence the slight difference in color
	from plot to plot.  The theoretical profiles are shown for
	ages 44, 118 and 211 years with respect to the first frame in
	Fig.~\ref{fig:cycle15f}. The size of our computational grid
	corresponds with the field of view of the observational image
	($131\arcsec \times 131\arcsec$) and a distance of 120 pc.}
	\label{fig:coldensplots}
\end{figure*}
%--------------------------------------------------------------------------
%
To enable a qualitative comparison of our results with recent
observations of \object{IRC +10216} (Mauron \& Huggins 1999; 2000), we
have produced Fig.~\ref{fig:coldensplots}. The left frame is adapted
from \cite*{MauronHuggins:1999} (their Fig.~3). It shows the composite
$B + V$ image of \object{IRC +10216}, with an average radial profile
subtracted to enhance the contrast. We compare this image with the
dust column density as a function of radius for a number of snapshots
in our calculation. The size of our computational grid (extended to
1287 $R_{*}$) corresponds with the field of view of the observational
image (131'' $\times$ 131'') and a distance of 120 pc. We too, have
subtracted an average radial density profile to enhance the
contrast. Comparing dust column density to the observed intensity
makes sense, since in the optically thin limit, the observed
intensity, due to illumination by the interstellar radiation field is
proportional to the column density along any line of sight
(\cite{MauronHuggins:2000}). We used the results of the 1.5--fluid
computation to produce Fig.~\ref{fig:coldensplots} because there the
short time scale structures are visible, whereas they are suppressed
in the two--fluid model because the latter isn't always second order
accurate.  Note that the fact that in our calculated images all shells
appear to be perfectly round is simply due to our assumption of
spherical symmetry. The two dimensional plots were produced by simply
rotating the spherical symmetric profile. In view of the fact that our
calculations indicate that the chemical--dynamic system that regulates
the behavior of the envelope is extremely stiff and reacts violently
to all kinds of changes, we think that it is rather unlikely that the
observed circumstellar shells are indeed complete. It is intriguing to
see that this idea is supported by the recent observations by
\cite*{MauronHuggins:2000}, which show that most shells, although they
may extend over much larger angles at lower levels, are prominent over
about $45\degr$. \\
As was mentioned before, Fig.~\ref{fig:coldensplots} only offers a
qualitative comparison with the observations. It can, however be used
to establish that the spacing of the shells, small scale structure
inside, large scale structure outside, is similar in the observations
and calculations. This, is not surprising however, since merging of
shells of various widths is due to dispersion, as was mentioned in
Section \ref{sec:resultsubsection}.
\subsection{The timescale of mass loss variations}
\label{sec:pulsationtimescale}
The characteristic time scale of the variability corresponds to {\em
the time needed by the rarefaction wave to cross the region between
the sonic point and the innermost point of the nucleation zone}. The
width of this region is, depending on the phase, a few times $10^{14}$
to $10^{15}$ cm. The velocity of the rarefaction wave equals the gas
velocity minus the local sound velocity and is typically a few times
$10^{4}$ to $10^{5}$ cm s$^{-1}$, also depending on the phase of the
variability. The resulting time scale is roughly 50 to 500 years,
which indeed corresponds to the time separation between two maxima in
the mass loss rate in our calculation.
\subsection{Discussion}
We found that the fact that the average grain size reacts strongly to
the density structure is an essential ingredient for the formation of
variability in the outflow. This explains why
\cite*{Mastrodemosea:1996} and \cite*{SteffenSchoenberner:2000}, who
also performed time dependent, two--fluid computations, but did not
take into account self consistent grain growth, did not encounter mass
loss variations in the outflow.\\
Also, grain drift occurs to be essential for variations in the mass
loss rate. If grains can drift with respect to the gas, they can form
regions of higher (or lower) density and/or size independently from
the gas.\\
Periodic variability in the mass loss rate occurs in both the
1.5--fluid and the two--fluid calculations, because grains are allowed
to drift in both cases.  Both calculations give somewhat different
results, though.  Probably, assuming equilibrium drift a priori, as
was done in the 1.5--fluid computation, influences the results, even
if the grains in the two--fluid model turn out to drift at the
equilibrium drift velocity as well. There are two reasons for this.
First, the fact that equilibrium drift has established itself at the
end of a numerical time step, does not mean that there has been
equilibrium always during this specific time step.  Hence, integration
of the drag force over the time step provides a better value of the
momentum transfer than multiplication of the drag force with the
duration of the time step, c.f. Appendix \ref{sec:append}. Second, the
value of the equilibrium drift velocity in the 1.5--fluid calculation
is indirectly determined by the dynamics, whereas in the two--fluid
case there is a direct influence. Also, the fact that the 1.5-fluid
calculation is second order accurate, but in the two-fluid calculation
this level of accuracy is not always achieved, will lead to
differences in the results.\\
We have not taken into account radiative transfer to solve the energy
structure in the envelope. Also, we used a grey absorption coefficient
in the radiative force and we did not calculate the grain temperature.
These are severe limitations of the model. However, we believe that
they do not influence the general conclusion that dynamics and
chemistry together can lead to time dependent structures.  It is more
likely that taking into account the temperature structure determined
by the optical properties of the grain population will make the
variability even more pronounced. This is inferred from previous
calculations by \cite*{Fleischerea:1992} in which the interaction
between atmospheric dynamics and radiative transfer was solved,
imposing a time dependent inner boundary. Recently,
\cite*{Wintersea:2000} performed similar calculations, also without
the piston at the inner boundary. Their results also indicate that the
coupling between the sensitive grain chemistry and the dynamics can
lead to variability in the wind.\\
The role of the inner boundary in calculations as presented here is
extremely important.  It is possible to generate wind variability
using a time dependent inner boundary. We did not do this: the inner
boundary that we have used was created to have as little influence on
the results as possible. It consists of a fixed advective flux which
can be modified by a diffusion term. The diffusive contribution to the
flux is proportional to the gradients of the flow variables near the
inner boundary, i.e. it is not externally prescribed. This is a
realistic approach, since the inner boundary is located in the
subsonic regime, where communication with lower layers is still
possible. In this respect a completely fixed inner boundary would be
less realistic.\\
We have referred to the quasi--periodic structure in our models as
``shells''. In order to prove that the structure is truly created in
the form of spherical shells one should perform three dimensional
hydrodynamics. Higher dimensionality will be a topic of future
research.
Shell structure is observed around only a small number of Post--AGB
objects and PNe. It is possible that the majority of objects doesn't
have shells. A stationary wind can definitely exist if for some reason
the equilibrium drift velocity is relatively low. This can be the case
if the luminosity of the star is low. This will limit the mutual
motion of both fluids and hence the value of the gas to dust density
ratio so that the outflow will remain more smooth.
%
%----------------------------------------------------------------------------
%
\section{Conclusion}
Our calculations suggest that the sensitive interplay of grain
nucleation and dynamics, in particular grain drift, leads to
quasi--periodic winds on the AGB. The characteristic time scale for
the variability corresponds to the crossing of the subsonic nucleation
zone by the rarefaction wave.  This time scale also matches recent
observations of \object{IRC +10216}.\\
More generally, we would like to stress that two--fluid hydrodynamics
is important in order to reach self--consistency of the modeling
method since the validity of the assumption of equilibrium drift is
hard to check. {\em If} equilibrium drift is applied, it should be
calculated by demanding the grains and the gas to be equally
accelerated, rather than by equating the drag force and the radiation
pressure on grains, because grains do have mass.\\ 
Observations also imply that gas and grains may not be spatially
coupled (\cite{sylvesterea:1999}) and that variations in the gas to
dust ratio in the outflow may arise (\cite{Omontea:1999}).
%
%----------------------------------------------------------------------------
%
\begin{acknowledgements}
We thank Jan Martin Winters for providing us with the initial
stationary profile for \object{IRC +10216} and Garrelt Mellema for
carefully reading the manuscript. Furthermore, the authors wish to
thank the referee for reading the manuscript with great attention and
providing many constructive comments and critical remarks.
\end{acknowledgements}
%
%----------------------------------------------------------------------------
%
\appendix
\section{Calculating the drag force}
\label{sec:append}
\begin{figure}
	\resizebox{\hsize}{!}{\includegraphics{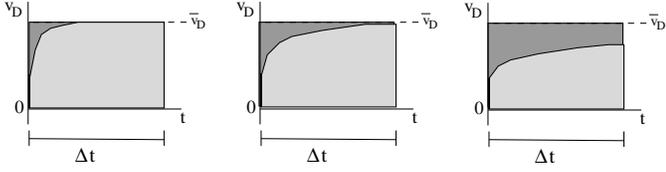}}
	\caption{Evolution towards equilibrium drift within a single
	time step: because it may take some time to establish
	equilibrium, calculating the momentum transfer by simply
	multiplying the drag force (which is proportional to
	$v_\mathrm{D}^2$) with $\Delta t$ overestimates the momentum
	transfer. The difference between the exact calculation and the
	equilibrium calculation increases with the time required to
	establish equilibrium drift and is represented by the
	dark color in the figures.}  \label{fig:vDtimesdt}
\end{figure}

To derive an expression for the drag force, we need to know about the
time evolution of the drift velocity. The gas--grain system will
always evolve towards a state in which grains drift at the equilibrium
drift speed, hence in which gas and grains undergo the same
acceleration. If, or how rapidly this state is reached depends on the
time needed to establish the equilibrium relative to the dynamical
time scale.  If one assumes that equilibrium drift is always valid,
the momentum transfer in a numerical time step can simply be
calculated by using the equilibrium value of the drift velocity in
Eq.(\ref{eq:fdragsimple}) and multiplying the drag force by the
duration of the time step. However, if, during a fraction of the
numerical time step, the drift velocity is lower than the equilibrium
value, assuming equilibrium drift when calculating the drag force will
overestimate the momentum transfer. This is illustrated in
Fig.~\ref{fig:vDtimesdt}. Although the error for a single time step
may be very small, the implications may be large for the time
dependent calculation. Note that, when assuming equilibrium drift, one
fixes the value of the drift velocity so that the gas and the dust
velocities are no longer independent flow variables. Therefore, when
calculating the momentum transfer assuming equilibrium drift one is
forced to do a 1.5-fluid calculation rather than a full two-fluid
calculation.
%
%------------------------------------------- fig:vDtimesdt=simis10276_f8
%--------------------------------------------------------------------------
\noindent
We will, hereafter, derive an expression for the time evolution of the
drift velocity. With this expression we can calculate the momentum
transfer as the integral of the drag force over the numerical time
step. No assumptions about the final drift velocity need to be made
and the derived expression can be used in a full two-fluid
calculation.\\
It is important to note that even if we find equilibrium drift in the
two component calculation this does not imply that it would have been
justified to assume equilibrium drift a priori. This can be seen from
Fig.~\ref{fig:vDtimesdt}. In both the first and the second panel
equilibrium drift is established within the duration of the numerical
time step, $\Delta t$, i.e., in both cases the output of the
hydrodynamics indicates equilibrium drift. Assuming equilibrium drift
throughout the time step would however only slightly overestimate the
momentum transfer in the first panel whereas is the second panel the
difference between the exact integral of the drag force over the time
step and equilibrium approximation would be much bigger.
\subsection{An analytical expression for the momentum transfer rate}
\label{sec:analyticmomtrans}
\begin{figure}
	\resizebox{\hsize}{!}{\includegraphics{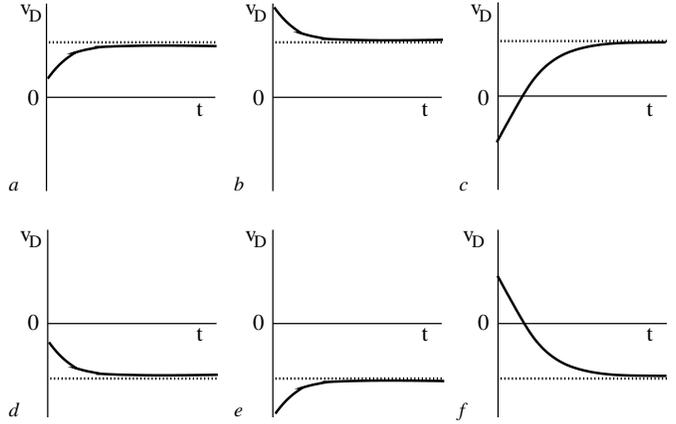}}
	\caption{Evolution towards equilibrium drift for various
	initial drift velocities. Upper panels: $g_\mathrm{D,tot} > 0
	\rightarrow \bar{v}_\mathrm{D} > 0$, lower panels: $g_\mathrm{D,tot} <
	0 \rightarrow \bar{v}_\mathrm{D} < 0$} 
	\label{fig:v0_and_vD}
\end{figure}

In this section we will derive an expression for the time evolution of
the drift velocity. Using this expression we can calculate the rate at
which momentum is transfered from grains to gas.\\
Fig.~\ref{fig:v0_and_vD} shows the six possible cases for reaching
equilibrium drift.  Note that both the initial drift and the
equilibrium value can be negative if the grains are less accelerated
than the gas.
%
%---------------------------------------------- fig:v0_and_vD=simis10276_f9
%--------------------------------------------------------------------------
%
We assume that the gas--grain interactions are completely
inelastic. Furthermore, we assume that after a collision with a grain,
a gas particle shares the acquired momentum with the surrounding gas
instantaneously (thermalization). This is realistic, since the mean
free path of gas--gas collisions is very small compared to the mean
free path for gas--grain encounters. We will not take into account
thermal motion because this enables us to derive an analytic
expression for the drag force. This will result in a somewhat lower
momentum transfer in the subsonic region. Farther out, the drift
velocity of the grains will dominate the collision rate anyway.\\
First, consider the motion of an individual gas particle between two
subsequent collisions with a grain:
\begin{equation}
v_\mathrm{g} \rightarrow v_\mathrm{g} + g_\mathrm{g,tot} \delta t + 
\frac{n_\mathrm{d}}{n_\mathrm{g}} \frac{\Delta p}{m_\mathrm{g}}
\end{equation}
Here, $v_\mathrm{g}$ is the velocity of the particle, after the
previous collision, $g_\mathrm{g,tot}$ is the total acceleration due
to gravity and the pressure gradient (but not the drag force), $\delta
t$ is the time interval between two collisions. The last term
represents the increase in the velocity as a result of the encounter
with the grain, and the (instantaneous) redistribution of the momentum
amongst the gas. $\Delta p$ is the amount of momentum transferred in a
single gas--grain collision,
\begin{equation}
\Delta p = \frac{m_\mathrm{g} m_\mathrm{d}}{m_\mathrm{g} + m_\mathrm{d}} u_\mathrm{D}
\label{eq:deltap}
\end{equation}
where $u_\mathrm{D}$ is the velocity of a grain with respect to the
gas immediately before the collision, $m_\mathrm{g,d}$ are the masses
of a gas particle (i.e. the mean molecular weight) and the (average)
grain mass.  A similar equation for the dust grain is
\begin{equation}
v_\mathrm{d} \rightarrow v_\mathrm{d} + g_\mathrm{d,tot} \delta t - 
\frac{\Delta p}{m_\mathrm{d}}
\end{equation}
The drift velocity after a collision, $v_\mathrm{D}$, can now be
expressed in terms of the drift velocity immediately before the
encounter, $u_\mathrm{D}$, as follows:
\begin{equation}
v_\mathrm{D} = \Omega u_\mathrm{D}
\label{eq:reluDandvD}
\end{equation}
in which
\begin{equation}
\Omega = \frac{\rho_\mathrm{g} m_\mathrm{d} - \rho_\mathrm{d} m_\mathrm{g}}
              {\rho_\mathrm{g}(m_\mathrm{g} + m_\mathrm{d})}
\end{equation}
\begin{equation}
u_\mathrm{D} = u_\mathrm{d} - u_\mathrm{g} 
	 = v_\mathrm{d} - v_\mathrm{g} + (g_\mathrm{d,tot} - g_\mathrm{g,tot}) \delta t
\label{eq:uDequd_ug}
\end{equation}
In the following, we will write $g_\mathrm{D,tot}$ for the relative
acceleration, $g_\mathrm{d,tot} - g_\mathrm{g,tot}$. The ``mean free
travel time'', $\delta t$, of a grain can be found by solving the
quadratic equation for the mean free path, $\lambda$, of a grain
\begin{equation}
\lambda = v_\mathrm{D} \delta t + \frac{1}{2} g_\mathrm{D,tot} \delta t^2
\label{eq:lambdaeq}
\end{equation}
Note that the mean free path can become negative if the initial drift
velocity, $v_\mathrm{D}$, and/or the relative acceleration
$g_\mathrm{D,tot}$ is negative.  If grains are not significantly
accelerated between two subsequent collisions with gas particles,
i.e. if $v_\mathrm{D} \gg g_\mathrm{D,tot} \delta t$,
Eq.(\ref{eq:lambdaeq}) simply becomes
\begin{equation}
\lambda = v_\mathrm{D} \delta t
\label{eq:lambda_v}
\end{equation}
so that $\delta t = \lambda/v_\mathrm{D}$. On the other hand, if the
acceleration of a grain between two collisions is so large that its
initial (drift) velocity is negligible, Eq.(\ref{eq:lambdaeq}) reads
\begin{equation}
\lambda = \frac{1}{2}g_\mathrm{D,tot}\delta t^2
\label{eq:lambda_gt}
\end{equation}
and $\delta t = \sqrt{2\lambda/g_\mathrm{D,tot}}$. The boundary
between the two regimes lies at the drift velocity for which $2
v_\mathrm{D} = g_\mathrm{D,tot}\delta t$. With $\delta t$ given by the
solution of Eq.(\ref{eq:lambdaeq}) we find that if
\begin{equation}
|v_\mathrm{D}| < \frac{1}{2}\sqrt{\lambda g_\mathrm{D,tot}}
\label{eq:lambda_zone}
\end{equation}
Eq.(\ref{eq:lambda_gt}) can be used instead of Eq.(\ref{eq:lambdaeq}).
In the current context of dust forming stellar winds, the quantity
$\Omega$ will always be nearly equal to unity\footnote{E.g. for a
typical dust to gas mass ratio $\rho_\mathrm{d}/\rho_\mathrm{g}= 1.0
\times 10^{-2}$ and for grains consisting of $10^{10}$ momomers
($m_\mathrm{d}/m_\mathrm{g}= 1.0 \times 10^{10}$) we find
$\Omega\approx 1 - 10^{-10}$}, so that $\bar{v}_\mathrm{D} \gg
\frac{1}{2}\sqrt{\lambda g_\mathrm{D,tot}}$. Hence, the zone in
velocity space where grain acceleration is significant is extremely
narrow. If the drift velocity is zero at some time (see
e.g. Fig.~\ref{fig:v0_and_vD}.c,f), it follows from
Eq.(\ref{eq:reluDandvD}), (\ref{eq:uDequd_ug}) and
(\ref{eq:lambda_gt}) that the drift velocity will be larger than
$\frac{1}{2} \sqrt{\lambda g_\mathrm{D,tot}}$ after a single collision
unless $\Omega < 1/\sqrt{8}$. This implies that we can safely apply
Eq.(\ref{eq:lambda_v}) for all values of $v_\mathrm{D}$.\\
In the following we will present a method to derive an expression for
the momentum transfer, which applies to all possible scenarios (see
Fig.~\ref{fig:v0_and_vD}) to reach equilibrium drift. We limit
ourselves to the derivation for the case $g_\mathrm{D,tot} > 0$
(Fig.~\ref{fig:v0_and_vD}.a,b,c), the derivation for negative
acceleration is analogous.\\
Application of Eq.(\ref{eq:lambda_v}) and Eq.(\ref{eq:uDequd_ug}) in
Eq.(\ref{eq:reluDandvD}) gives rise directly to a recurrence relation
for $v_\mathrm{D}$:
\begin{equation}
v_\mathrm{D}(t_{i+1}) = \Omega u_\mathrm{D}(t_{i+1}) = 
		    \Omega \left(v_\mathrm{D}(t_i) + 
		    \frac{g_\mathrm{D,tot} \lambda}{v_\mathrm{D}(t_i)}\right)
\label{eq:vDrecurlindrift}
\end{equation}
From this, and $\delta t$ given by Eq.(\ref{eq:lambda_v}), a
differential equation for the drift velocity as a function of time can
be derived:
\begin{equation}
\frac{\Delta v_\mathrm{D}}{\Delta t} \simeq 
\frac{\mathrm{d}  v_\mathrm{D}}{\mathrm{d}  t} = \frac{\Omega -1}{\lambda}v_\mathrm{D}^2 + 
\Omega g_\mathrm{D,tot}
\label{eq:diffeqlindrift}
\end{equation}
This equation can be easily solved for $t(v_\mathrm{D})$,
\begin{eqnarray}
t(v_\mathrm{D}) &=& \frac{\lambda}{\sqrt{\Omega(\Omega-1) g \lambda}} \nonumber\\
&\Biggl[&\arctan \left( \frac{(\Omega-1)v_\mathrm{D}(t)}{\sqrt{\Omega(\Omega-1) g \lambda}} \right) \Biggr. - 
\nonumber\\
&\Biggl.& \arctan \left( \frac{(\Omega-1)v_\mathrm{D}(0)}{\sqrt{\Omega(\Omega-1) g \lambda}} \right) \, \Biggr]
\label{eq:t_Lgoppositesign}
\end{eqnarray}
where $g$ stands for $g_\mathrm{D,tot}$.\\
First, consider the case where $v_\mathrm{D}(0)> 0$ (and $g > 0$). In
this case the mean free path $\lambda$ will always be positive.
Because $\Omega$ is always smaller than unity and $\lambda$ and $g$
have equal signs this is rewritten as
\begin{eqnarray}
t(v_\mathrm{D}) &=& \frac{\lambda}{\sqrt{\Omega(1-\Omega) g \lambda}} \nonumber\\ 
&\Biggl[& \mbox{arctanh}\left(\frac{(1-\Omega)v_\mathrm{D}(t)}{\sqrt{\Omega(1-\Omega) g \lambda}}\right) - \Biggr. \nonumber\\
&\Biggl.& \mbox{arctanh}\left(\frac{(1-\Omega)v_\mathrm{D}(0)}{\sqrt{\Omega(1-\Omega) g \lambda}}\right) \, \Biggr]
\end{eqnarray}
This expression can be simplified by realizing that from
Eq.(\ref{eq:vDrecurlindrift}) it follows that the equilibrium drift
velocity is given by
\begin{equation}
\bar{v}_\mathrm{D} = \sqrt{\frac{\Omega}{1-\Omega} \lambda g}
\label{eq:vdriftlimitpos}
\end{equation}
and that the equilibration time scale is
\begin{equation}
\tau_\mathrm{eq} = \frac{1}{\sqrt{\Omega(1-\Omega)g/\lambda}}
\label{eq:taueq}
\end{equation}
so that
\begin{eqnarray}
t(v_\mathrm{D}) = \tau_\mathrm{eq} 
&&\left[ \mbox{arctanh}\left(\frac{v_\mathrm{D}(t)}{\bar{v}_\mathrm{D}}\right) - \right.\nonumber\\
&&\left. \,\,\, \mbox{arctanh}\left(\frac{v_\mathrm{D}(0)}{\bar{v}_\mathrm{D}}\right) \right]\\
            = \tau_\mathrm{eq}\
&\, & \,\,\, \mbox{arctanh}\left(\frac{(v_\mathrm{D}(t)-v_\mathrm{D}(0))\bar{v}_\mathrm{D}}
	                 {\bar{v}_\mathrm{D}^2 - v_\mathrm{D}(t) v_\mathrm{D}(0)}\right)
\end{eqnarray}
Note that addition of the $\mbox{arctanh}$ terms causes the expression
to be valid for initial values $v_\mathrm{D}(0) > \bar{v}_\mathrm{D}$
(see Fig.~\ref{fig:v0_and_vD}.b) as well.  Inversion leads to an
expression for the drift velocity as a function of time:
\begin{equation}
v_\mathrm{D}(t) = \bar{v}_\mathrm{D} \frac{v_\mathrm{D}(0) + \bar{v}_\mathrm{D} \Theta(t)}
				  {\bar{v}_\mathrm{D} + v_\mathrm{D}(0) \Theta(t)}
\end{equation}
with
\begin{equation}
\Theta(t) =  \mbox{tanh} ( t/\tau_\mathrm{eq} )
\end{equation}
The drag force (density) is the product of the number of gas--grain
collisions per unit volume and time and the momentum transfer per
collision.  In Eq.(\ref{eq:fdragsimple}), the amount of momentum
transfer in a single collision was simply assumed to be
$m_\mathrm{g}v_\mathrm{D}$, now we use the more accurate form for
$\Delta p$ which follows from Eqs.(\ref{eq:deltap}),
(\ref{eq:reluDandvD}), (\ref{eq:lambda_v}).  With $\lambda =
1/\Sigma_\mathrm{d}n_\mathrm{g}$ we then find
\begin{equation}
f_\mathrm{drag} = \Sigma_\mathrm{d} \rho_\mathrm{g} \frac{n_\mathrm{g} n_\mathrm{d}}{n_\mathrm{g} - n_\mathrm{d}} 
|v_\mathrm{D}| v_\mathrm{D}
\label{eq:fdrag}
\end{equation}
The standard way to calculate the amount of momentum transfer per
numerical time step is simply multiplying the drag force with the
duration of the time step. Now that we have derived an expression for
the drift velocity as a function of time we can calculate the momentum
transfer more accurate, by integrating Eq.(\ref{eq:fdrag}), assuming
$n_\mathrm{g,d},m_\mathrm{g,d}$ are constant:
\begin{eqnarray}
&&\int_{0}^{\tau}f_\mathrm{drag} \mbox{d}t 
=  \Sigma_\mathrm{d} \rho_\mathrm{g} \frac{n_\mathrm{g} n_\mathrm{d}}{n_\mathrm{g} - n_\mathrm{d}}
    \tau_\mathrm{eq} \bar{v}_\mathrm{D}^2 \Biggl[\frac{\tau}{\tau_\mathrm{eq}} + \nonumber \\ 
	 &&\left(\frac{v_\mathrm{D}(0)}{\bar{v}_\mathrm{D}} - 
	          \frac{\bar{v}_\mathrm{D}}{v_\mathrm{D}(0)}\right) 
            \left( \frac{v_\mathrm{D}(0) \tanh(\tau / \tau_\mathrm{eq})}
	                {v_\mathrm{D}(0)\tanh(\tau / \tau_\mathrm{eq}) + \bar{v}_\mathrm{D}}\right)\Biggr]
\label{eq:fdraginttot_a}
\end{eqnarray}
If the initial drift velocity and the total acceleration have opposite
sign ($v_\mathrm{D}(0)< 0, g > 0$, see Fig.~\ref{fig:v0_and_vD}.c) the
integral representing the total momentum transfer is split into two
parts,
\begin{equation}
\int_{0}^{\tau} f_\mathrm{drag} \mbox{d}t = \int_{0}^{t(v_D=0)} f_\mathrm{drag} \mbox{d}t +
\int_{t(v_D=0)}^{\tau} f_\mathrm{drag} \mbox{d}t 
\label{eq:fdragintsplit}
\end{equation}
where $t(v_\mathrm{D}=0)$ follows from Eq.(\ref{eq:t_Lgoppositesign}):
\begin{eqnarray}
t(v_\mathrm{D}=0) &=& \frac{-\lambda}{\sqrt{\Omega(\Omega-1) g \lambda}} \nonumber \\
&&\arctan \left( \frac{(\Omega-1)v_\mathrm{D}(0)}{\sqrt{\Omega(\Omega-1) g \lambda}}\right)
\label{eq:t_vDiszero}
\end{eqnarray}
Note that the mean free path of a grain, $\lambda$, is negative as
long as the drift velocity is negative.  The second term in
Eq.(\ref{eq:fdragintsplit}) is calculated as in the case
$v_\mathrm{D}(0)> 0$, simply taking $v_\mathrm{D}(0)= 0$. In order to
compute the first term, Eq.(\ref{eq:t_Lgoppositesign}) is inverted. We
find
\begin{equation}
v_\mathrm{D}(t) = \bar{v}_\mathrm{D} \frac{v_\mathrm{D}(0) + \bar{v}_\mathrm{D} \Theta'(t)}
				  {\bar{v}_\mathrm{D} - v_\mathrm{D}(0) \Theta'(t)}
\end{equation}
in which
\begin{equation}
\Theta'(t) =  \mbox{tan} ( t/\tau'_\mathrm{eq})
\end{equation}
\begin{equation}
\bar{v}_\mathrm{D} = \sqrt{\frac{\Omega}{\Omega-1} \lambda g}
\label{eq:vdriftlimitneg}
\end{equation}
\begin{equation}
\tau'_\mathrm{eq} =  \frac{1}{\sqrt{\Omega(\Omega-1)g / \lambda}}
\end{equation}
Inserting this into Eq.(\ref{eq:fdrag}) and integrating over the
interval $t=0,t(v_D=0)$, we obtain
\begin{eqnarray}
\int_{0}^{t(v_D=0)} f_\mathrm{drag} \mbox{d}t 
&=& -\Sigma_\mathrm{d} \rho_\mathrm{g} \frac{n_\mathrm{g} n_\mathrm{d}}{n_\mathrm{g} - n_\mathrm{d}} 
     \tau'_\mathrm{eq} \bar{v}_\mathrm{D}^2\nonumber\\
    &&\left[ -\frac{v_\mathrm{D}(0)}{\bar{v}_\mathrm{D}} + 
             \arctan\left(\frac{v_\mathrm{D}(0)}{\bar{v}_\mathrm{D}}\right)\right]
\end{eqnarray}
Note that the minus sign accounts for the fact that the momentum
transfer contains an integral over $|v_\mathrm{D}|v_\mathrm{D}$ rather
than an integral over $v_\mathrm{D}^2$.  Finally, for the complete
integral, Eq.(\ref{eq:fdragintsplit}), we find
\begin{eqnarray}
&&\int_{0}^{\tau} f_\mathrm{drag} \mbox{d}t = 
\Sigma_\mathrm{d} \rho_\mathrm{g} \frac{n_\mathrm{g} n_\mathrm{d}}{n_\mathrm{g} - n_\mathrm{d}}
\tau_\mathrm{eq} \bar{v}_\mathrm{D}^2\nonumber\\
&&\left[\frac{\tau}{\tau_\mathrm{eq}} - 
\mbox{tanh} \left(\frac{\tau}{\tau_\mathrm{eq}}  + 
                   \mbox{arctan}\left( \frac{v_\mathrm{D}(0)}{\bar{v}_\mathrm{D}} \right) \right) + 
 \frac{v_\mathrm{D}(0)}{\bar{v}_\mathrm{D}} \right]
\label{eq:fdraginttot_c}
\end{eqnarray}
As was to be expected Eq.(\ref{eq:fdraginttot_a}) and
Eq.(\ref{eq:fdraginttot_c}) are equal if $v_\mathrm{D}(0)= 0$.\\
Similar expressions for the total momentum transfer can be calculated
in the case of negative total acceleration (see
Fig.~\ref{fig:v0_and_vD}.d,e,f).\\ The above formulations for the
momentum transfer, in which no assumptions about the value of the
drift velocity or the completeness of momentum coupling have been
made, can be used as source terms in the momentum equations.
\subsection{Calculation of the equilibrium drift velocity}
\label{sec:eqdrift}
We have used the terms equilibrium drift velocity and limiting
velocity as equivalent. Here, we will show that both are indeed the
same. We equate the acceleration of the gas and the dust, rather than
equating the drag force and the radiation pressure of grains. In the
latter case one implicitly assumes that grains do not have mass
whereas the former leads to a general expression for the equilibrium
drift velocity.\\
From the equation of motion of a gas element,
\begin{equation}
\frac{\mathrm{d} v_\mathrm{g}}{\mathrm{d} t} = g_\mathrm{g,tot} + \frac{f_\mathrm{drag}}{\rho_\mathrm{g}}
\end{equation}
and its counterpart for a grain,
\begin{equation}
\frac{\mathrm{d} v_\mathrm{d}}{\mathrm{d} t} = g_\mathrm{d,tot} - \frac{f_\mathrm{drag}}{\rho_\mathrm{d}}
\end{equation}
we find that grains and gas are equally accelerated, and hence the
drift velocity has reached its equilibrium value, if
\begin{equation}
g_\mathrm{D,tot} = \frac{\rho_\mathrm{d} + \rho_\mathrm{g}}{\rho_\mathrm{d}\rho_\mathrm{g}} f_\mathrm{drag}
\end{equation}
With Eq.(\ref{eq:fdrag}), the equilibrium drift velocity is
\begin{equation}
\bar{v}_\mathrm{D} = \sqrt{ \frac{m_\mathrm{d} (n_\mathrm{g} - n_\mathrm{d})}{\rho_\mathrm{d} + \rho_\mathrm{g}} 
g_\mathrm{D,tot} \lambda}
\label{eq:vdrifteq}
\end{equation}
Thus, we have now derived an expression for the equilibrium drift
velocity without having to assume complete momentum coupling.  This
expression is indeed the same as Eq.(\ref{eq:vdriftlimitpos}), which
represents the limiting drift velocity.
%
%----------------------------------------------------------------------------
%
\bibliographystyle{aabib}

%
%--------------------------------------------------------------------------
%
\end{document}